\documentclass[aps,11pt,amssymb,amsmath]{revtex4}
\usepackage{amsthm}
\usepackage{dsfont}
\usepackage{bm}
\usepackage{enumerate}
\usepackage[all]{xy}

\begin{document}

\newcommand{\new}[1]{\framebox{$\blacktriangleright$\textbf{#1}$\blacktriangleleft$}}
\newcommand{\newm}[1]{\blacktriangleright#1\blacktriangleleft}


\theoremstyle{remark}
\newtheorem{rema}{Remark}[section]
\theoremstyle{plain}
\newtheorem{prop}[rema]{Proposition}
\theoremstyle{definition}
\newtheorem{defi}[rema]{Definition}

\title{Quantum manifolds with classical limit}

\author{Manuel Hohmann}
\author{Raffaele Punzi}
\author{Mattias N.\,R. Wohlfarth}
\affiliation{Zentrum f\"ur Mathematische Physik und II. Institut f\"ur Theoretische Physik, Universit\"at Hamburg, Luruper Chaussee 149, 22761 Hamburg, Germany}

\begin{abstract}
We propose a mathematical model of quantum spacetime as an infinite-dimensional manifold locally homeomorphic to an appropriate Schwartz space. This extends and unifies both the standard function space construction of quantum mechanics and the manifold structure of spacetime. In this picture we demonstrate that classical spacetime emerges as a finite-dimensional manifold through the topological identification of all quantum points with identical position expectation value. We speculate on the possible relevance of this geometry to quantum field theory and gravity.
\end{abstract}

\maketitle


\section{Motivation}
Current research on quantum gravity can be classified into essentially two different approaches; the first is {\it quantization}, the second could be termed {\it quantum construction}.

Perhaps not surprisingly, the two main contenders at present for a full theory of quantum gravity, namely string theory and loop quantum gravity, both are based on the quantization of a well-understood classical theory. As is well-known, gravity  comes into string theory through the requirement that the conformal invariance of the classical string also hold in the quantum theory, which yields various supergravity theories as low energy effective descriptions of quantum strings~\cite{Polchinski}. From the geometric point of view these metric theories easily recover the classical manifold. However, a number of questions about the geometric interpretation of the background fields besides the metric, such as the dilaton and Neveu-Schwarz two form, remain open. A lot of work recently goes towards understanding the role of these generalized backgrounds~\cite{Grana:2008yw,Hull:2007zu,Punzi:2007di,Zabzine:2006uz,Schuller:2005ru,Gualtieri:2003dx}.
Loop quantum gravity on the other hand is based on canonical quantization of Einstein's general relativity in Ashtekar variables~\cite{Rovelli, Thiemann:2001yy}. A number of promising results like the quantization of the area and volume operators has been obtained in this framework, and applied, for example, to the discussion of the big bang in the very early universe~\cite{Ashtekar:2003hd,Bojowald:2001xe}. However, the geometric interpretation of loop quantum gravity is still largely unresolved; although the theory originates from a metric classical background, it so far lacks the re-emergence of the manifold geometry from the spin network states in a suitable classical limit~\cite{Nicolai:2005mc}.

The main idea behind quantum construction is that the classical manifold picture of spacetime should be changed, even break down, at small length scales comparable to the Planck length, or at very high energy scales. Thus one starts out with quantum modifications of the basic background structures, in some sense reversing the procedure of quantization. A wide-ranging number of ideas fall into this category. On the phenomenological level many approaches consider the emergence of modified energy momentum dispersion relations~\cite{AmelinoCamelia:2008qg,Girelli:2006fw}. A geometrical basis for such an assumption, however, is rarely given~\cite{Rovelli:2008cj}. On the rigorous mathematical side there are various ideas for the discretization of spacetime, such as the description of spacetime as a causal set~\cite{Sverdlov:2008sx,Bombelli:1987aa} or via dynamical triangulations~\cite{Ambjorn:2005qt}, or for the short-distance modification of spacetime for instance by non-commutative geometry~\cite{Chamseddine:2007hz,Connes:1996gi}. Some of the mathematically rigorous approaches are not easily interpreted in terms of a smooth classical limit manifold.

\begin{figure}[ht]
\begin{equation*}
\xymatrix{
M \ar@{<~>}[dr]_{\text{local homeomorphism}} & & \mathcal{S} \ar[dl]^{\text{position expectation}}\\
& \mathbb{R}^n &}
\end{equation*}
\caption{\label{diagram1}\textit{Manifold geometry $M$ and quantum function spaces $\mathcal{S}$ limiting to $\mathbb{R}^n$.}}
\end{figure}
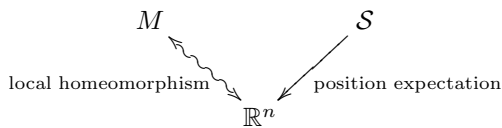

Our work in this article falls within the second category of quantum construction; we construct and explore a modified structure of spacetime, rather than quantizing a given classical theory. We are conservative in the sense that we use as main ingredients ideas already known from other work on the geometry of quantum theory; one such idea is that of infinite-dimensional manifolds~\cite{Kryukov:2007,Kryukov:2004mz,Cirelli:1990a,Cirelli:1990b,Kibble:1978tm}. However, we employ this idea in a very different way. More precisely, we propose a mathematical model of quantum spacetime based on an infinite-dimensional manifold locally homeomorphic to an appropriate space of Schwartz functions. The idea behind this is as follows. Consider  the space $\mathbb{R}^n$ as the simplest arena for classical mechanics, i.e., as the space of positions providing labels for classical physical events. This background appears as the limit of two more general constructions, as shown in the diagram in figure~\ref{diagram1}. The first is the manifold geometry at the heart of general relativity, where spacetime is locally homeomorphic to~$\mathbb{R}^n$; in this setting the idea of labels for events is taken seriously in the sense that different labellings can be chosen provided they are diffeomorphic. The second construction reducing to~$\mathbb{R}^n$ is that of quantum mechanics. Here observables become operators on a Hilbert space, or more accurately, on the Schwartz space~$\mathcal{S}(\mathbb{R}^n)$ of fast-decreasing functions over~$\mathbb{R}^n$, which is reobtained as the set of all possible position expectation values.

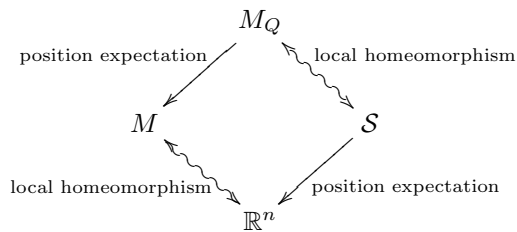
\begin{figure}[ht]
\begin{equation*}
\xymatrix{
& M_Q \ar@{<~>}[dr]^{\text{local homeomorphism}} \ar[dl]_{\text{position expectation}} &\\
M \ar@{<~>}[dr]_{\text{local homeomorphism}} & & \mathcal{S} \ar[dl]^{\text{position expectation}}\\
& \mathbb{R}^n &}
\end{equation*}
\caption{\label{diagram2}\textit{Quantum manifold $M_Q$ unifying manifold geometry~$M$ and quantum function spaces $\mathcal{S}$.}}
\end{figure}

The mathematical construction that we present below unifies both the manifold and function space ideas, which might be seen as the fundamental structures behind general relativity and quantum theory, respectively. We are able to complete the diagram above by introducing the geometry of a quantum manifold~$M_Q$ that is, on the one hand, locally homeomorphic to the Schwartz space~$\mathcal{S}(\mathbb{R}^n)$, and, on the other hand, allows the computation of position expectation values that recover the classical manifold. This idea is nicely expressed in the completed diagram of figure~\ref{diagram2}. We wish to emphasize that it will be a feature of the mathematical structure of the quantum manifold explicitly to avoid the problems of other quantum constructions with the classical limit.

To put our proposal into a bigger perspective, we might speculate that physics should be formulated as field theories over the quantum manifold; seen from the classical perspective these theories should thus at the same time be geometrically well-defined and show quantum behaviour. In other words, if this can be made precise, classical fields on the quantum geometry would become quantum fields on the classical geometry. Then this would also lead to an exciting new formulation of quantum gravity, simply as a classical geometric gravity theory on the infinite-dimensional quantum manifold, which becomes quantum from the point of view of the classical observer.

In this article we take the first important step in this program.  We transform the idea of the quantum manifold into a concrete, and mathematically precise, geometric concept, and explore some of its properties. Section~\ref{modelspace} discusses the Schwartz space on which we will model our quantum geometry. The infinite-dimensional quantum manifold is precisely defined in section~\ref{qmfd}. In section~\ref{cllimit} we prove the central result that a finite-dimensional differentiable manifold emerges as a classical limit by topological identification of functions with coinciding position expectation values. Moreover, we are able to show in section~\ref{fiberbundle} that any quantum manifold carries the structure of a fiber bundle over its associated classical limit. Section~\ref{trivq} demonstrates that any classical manifold can be trivially quantized, which ensures the existence of quantum manifolds. We end with a discussion in section~\ref{discuss}. Two appendices complete this article; appendix~\ref{topology} reviews the required topological notions, while appendix~\ref{proofs} proves some of the results stated in the main text.

\section{Model space}\label{modelspace}
We aim at a model for quantum spacetime that unifies key ingredients both from general relativity and quantum theory. As motivated in the introduction, the first ingredient we choose is the geometric structure of a manifold, the second is the notion that function spaces should play a major role. Thus we will construct a very specific infinite-dimensional manifold that is locally homeomorphic to a function space. Since we are interested in a quantum spacetime model with a classical limit we will also tie in the idea of position expectation values. In extension of standard quantum mechanics, we will find it appropriate for our needs to model the infinite-dimensional quantum manifold on the Schwartz space $\mathcal{S}(\mathbb{R}^n)$ of fast-decreasing functions over $\mathbb{R}^n$. In this section we will discuss this model space in more detail, as a necessary and important preparation for the definition of the quantum manifold in the next section. 

First consider the Hilbert space $L^2(\mathbb{R}^n)$ of square-integrable complex functions over $\mathbb{R}^n$ with the scalar product
\begin{equation}
\left<f,g\right>=\int_{\mathbb{R}^n}d\bm{x}\,f(\bm{x})^*g(\bm{x})
\end{equation}
between two functions $f$ and $g$. We canonically define the position operators $Q^i$ and the momentum operators $P_i$ for $i=1\dots n$; they act on functions $f$ as $(Q^i f)(\bm{x})=x^i f(\bm{x})$ and $(P_i f)(\bm{x})=-i\partial_i f(\bm{x})$. These operators are not defined on the whole of $L^2(\mathbb{R}^n)$ since their application may destroy square-integrability. They are, however, defined on the Schwartz space $\mathcal{S}(\mathbb{R}^n)$, which is a dense subset of~$L^2(\mathbb{R}^n)$ and closed under the operation of $\bm{Q}$ and $\bm{P}$:
\begin{equation}
\mathcal{S}(\mathbb{R}^n) = \Big\{f \in C^\infty(\mathbb{R}^n) \,|\, \forall\, \bm{\alpha},\bm{\beta} \in \mathbb{N}^n: \sup_{\bm{x} \in \mathbb{R}^n}|x^{\bm{\alpha}}D_{\bm{\beta}}f(\bm{x})| < \infty\Big\}.
\end{equation}
For later convenience we define $\mathcal{S}^{\neq 0}(\mathbb{R}^n)$ as the Schwartz space with the zero function removed,
\begin{equation}
\mathcal{S}^{\neq 0}(\mathbb{R}^n) = \mathcal{S}(\mathbb{R}^n)\setminus \{0\}\,.
\end{equation}
The Schwartz space, and by restriction also $\mathcal{S}^{\neq 0}(\mathbb{R}^n)$, are topological spaces. The family 
\begin{equation}
\|f\|_{\bm{\alpha},\bm{\beta}} = \sup_{\bm{x} \in \mathbb{R}^n}|x^{\bm{\alpha}}D_{\bm{\beta}}f(\bm{x})|
\end{equation}
for all multiindices $\bm{\alpha},\bm{\beta}$ is a family of seminorms which generates a topology on \(\mathcal{S}(\mathbb{R}^n)\). This topology is called the natural topology. Constructively, it is generated by the open balls $V^r_{\bm{\alpha},\bm{\beta}}(f_0)$ of radius $r>0$ around \(f_0 \in \mathcal{S}(\mathbb{R}^n)\),
\begin{equation}
V^r_{\bm{\alpha},\bm{\beta}}(f_0)=\{f \in \mathcal{S}(\mathbb{R}^n) \,|\, \|f_0 - f\|_{\bm{\alpha},\bm{\beta}} < r\}\,.
\end{equation}

We will now take a closer look on the respective topological duals $(L^2(\mathbb{R}^n))^*$ and $(\mathcal{S}(\mathbb{R}^n))^*$ of the Hilbert space and the Schwartz space, and their relations. By definition the topological dual is the space of linear functionals. Consider first the Hilbert space $L^2(\mathbb{R}^n)$. By the Riesz representation theorem the Hilbert space and its topological dual are isomorphic via the map $L^2(\mathbb{R}^n)\rightarrow (L^2(\mathbb{R}^n))^*,\,f\mapsto f^\dagger$ where $f^\dagger$ is defined by the scalar product as $f^\dagger(g)=\left<f,g\right>$. Now note that the Schwartz space is not a Hilbert space, but that we have a natural injection ${\mathcal{S}(\mathbb{R}^n) \hookrightarrow L^2(\mathbb{R}^n)}$. The dual of the Schwartz space is known as the space of tempered distributions. There exists a natural injection $(L^2(\mathbb{R}^n))^* \hookrightarrow (\mathcal{S}(\mathbb{R}^n))^*$; indeed, recalling that the elements of $(L^2(\mathbb{R}^n))^*$ are linear functionals on $L^2(\mathbb{R}^n)$, the injection is simply their restriction to~$\mathcal{S}(\mathbb{R}^n)$. The composition of the two injections and the isomorphism of $L^2(\mathbb{R}^n)$ onto its dual induces an anti-linear injection $\mathcal{S}(\mathbb{R}^n) \hookrightarrow (\mathcal{S}(\mathbb{R}^n))^*$. Working with the Schwartz space and its topological dual hence is nothing else than working with the Gelfand triple~\cite{Bohm:1989}, or rigged Hilbert space,
\begin{equation}
\mathcal{S}(\mathbb{R}^n) \,\subset\, L^2(\mathbb{R}^n) \cong (L^2(\mathbb{R}^n))^*\, \subset\, (\mathcal{S}(\mathbb{R}^n))^*\,.
\end{equation}

To link our construction below of the infinite-dimensional quantum manifold with the classical finite-dimensional picture of spacetime we will exploit the idea of position expectation values. As usual, the expectation value of a linear operator $O:\mathcal{S}\rightarrow\mathcal{S}$ is the map $\bar O:\mathcal{S}^{\neq 0}\rightarrow \mathbb{R}$ defined by
\begin{equation}
\bar O(f)=\frac{\left<f,Of\right>}{\left<f,f\right>}\,.
\end{equation}
Hence, for any $f\in\mathcal{S}^{\neq 0}$ the position expectation value is the vector $(\bar{\bm{Q}}(f))^i=\bar Q^i(f)$.

The position expectation value allows us to define a very special topology on $\mathcal{S}^{\neq 0}(\mathbb{R}^n)$, which we will call the expectation value topology. We will see below that this topology is essential in the definition of the quantum manifold and the recovery of classical spacetime. We take the expectation value topology to be the initial topology~$\iota(\mathcal{S}^{\neq 0}(\mathbb{R}^n),\bar{\bm{Q}})$ with respect to $\bar{\bm{Q}}$. The definition, see appendix~\ref{topology}, implies that the open sets of the expectation value topology are precisely of the form
\begin{equation}
\bar{\bm{Q}}^{-1}(W)=\left\{f\in\mathcal{S}^{\neq 0}\,|\,\bar{\bm{Q}}(f) \in W\right\},
\end{equation}
where $W\subset \mathbb{R}^n$ is open in the standard topology on $\mathbb{R}^n$. Thus this is the coarsest topology in which the function~$\bar{\bm{Q}}$ is continuous. 

Note that every set \(V \subset \mathcal{S}(\mathbb{R}^n)\) that is open in the initial topology is also open in the natural topology defined above. This is equivalent to the statement that \(\bar{\bm{Q}}\) is also continuous with respect to the natural topology on \(\mathcal{S}(\mathbb{R}^n)\). We will prove this statement in appendix~\ref{proofs}. There we will also prove that \(\bar{\bm{Q}}\) is not merely continuous but also differentiable with respect to the natural topology on \(\mathcal{S}(\mathbb{R}^n)\). Both these properties of \(\bar{\bm{Q}}\) will be needed as key ingredients when we construct the classical limit of a quantum manifold and show that it is in fact a differentiable manifold.

For practical computations, it is often convenient to use a different family of seminorms on the Schwartz space \(\mathcal{S}(\mathbb{R}^n)\). These seminorms are given by
\begin{equation}\label{pseminorms}
\|f\|_p = \left<f, (\bm{Q}^2 + \bm{P}^2 + \mathds{1})^pf\right>
\end{equation}
for all \(f \in \mathcal{S}(\mathbb{R}^n)\) and \(p \in \mathbb{N}\). They generate the so-called nuclear topology on \(\mathcal{S}(\mathbb{R}^n)\) via the open balls $V^r_p(f_0)$ of radius $r>0$ around \(f_0 \in \mathcal{S}(\mathbb{R}^n)\), 
\begin{equation}
V^r_p(f_0)=\{f \in \mathcal{S}(\mathbb{R}^n) | \|f_0 - f\|_p < r\}\,.
\end{equation}
It can be shown that the two families of seminorms \((\|f\|_p,\, p \in \mathbb{N})\) and \((\|f\|_{\bm{\alpha},\bm{\beta}},\, \bm{\alpha}, \bm{\beta} \in \mathbb{N}^n)\) generate the same topology~\cite{ReedSimon:1972}, i.e., sets are open in one topology if and only if they are in the other. Therefore, we may use the words nuclear topology and natural topology synonymously, and any statement valid for one of them will also be valid for the other.

\section{Quantum manifolds}\label{qmfd}
The Schwartz space of functions together with certain useful topologies has been discussed in some detail in the previous section. This space is to become the model space for the quantum manifold, and we are now in the position to present the central definitions of this article. Following the construction of classical manifolds in~\cite{Lang}, we will proceed to define the quantum manifold by first, equipping a basic set with an atlas, and second, by considering equivalence classes of compatible atlases.

We begin with the definition of a quantum atlas of dimension $n\in \mathbb{N}$ on a set $M_Q$.  The quantum atlas is a collection of pairs $\mathcal{A} = \{(U_i, \phi_i),\, i \in \mathcal{I}\}$ for some index set $\mathcal{I}$, called charts, which satisfy the following four conditions:
\begin{enumerate}[\it (i)]
\vspace{-6pt}\item Each $U_i$ is a subset of $M_Q$ and the $U_i$ cover $M_Q$.

\vspace{-6pt}\item Each $\phi_i$ is a bijection of $U_i$ onto a set $\phi_i(U_i) \subset \mathcal{S}^{\neq 0}(\mathbb{R}^n)$.

\vspace{-6pt}\item For each \(i, j\), the set \(\phi_i(U_i \cap U_j)\) is open in the expectation value topology $\iota(\mathcal{S}^{\neq 0}(\mathbb{R}^n),\bar{\bm{Q}})$.

\vspace{-6pt}\item For each \(i, j\), the transition map
\begin{equation}
\phi_{ji} = \phi_j \circ \phi_i^{-1}
\end{equation}
on the overlap of any two charts, $\phi_{ji}: \phi_i(U_i \cap U_j) \to \phi_j(U_i \cap U_j)$ is continuous in the expectation value topology and differentiable in the natural topology.
\end{enumerate}
From this definition, it is clear that the final quantum manifold will be a differentiable infinite-dimensional manifold locally homeomorphic to $\mathcal{S}^{\neq 0}(\mathbb{R}^n)$. In contrast to the usual definition of an atlas, two different topologies are introduced. We will see in the following section that this is central for the existence of a classical limit. Before we can state the definition of the quantum manifold, we need to consider equivalence classes of compatible atlases.

Two quantum atlases \(\mathcal{A}, \mathcal{A}'\) of common dimension \(n\) on a set \(M_Q\) are said to be compatible, if also their union \(\mathcal{A} \cup \mathcal{A}'\) is a quantum atlas of dimension \(n\) on \(M_Q\). Note that compatibility of quantum atlases so defined is an equivalence relation. Reflexivity and symmetry are immediate, so we only need to check that compatibility of atlases is transitive: given three atlases \(\mathcal{A}, \mathcal{A}', \mathcal{A}''\) with \(\mathcal{A} \cup \mathcal{A}'\) and \(\mathcal{A}' \cup \mathcal{A}''\) also being atlases, \(\mathcal{A} \cup \mathcal{A}''\) should also be an atlas. We now check the conditions from the definition in turn. Clearly, \(\mathcal{A} \cup \mathcal{A}''\) is a collection of pairs \((U_i, \phi_i)\), where the~\(U_i\) cover~\(M_Q\) and the \(\phi_i\) are bijections of \(U_i\) onto subsets of $\mathcal{S}^{\neq 0}(\mathbb{R}^n)$, which gives conditions~\textit{(i)} and~\textit{(ii)}. We still have to check the remaining two, namely, that for all \(i, j\), the sets  \(\phi_i(U_i \cap U_j)\) are open in the expectation value topology of $\mathcal{S}^{\neq 0}(\mathbb{R}^n)$ and the maps \(\phi_{ji} = \phi_j \circ \phi_i^{-1}\) are continuous in the expectation value topology and differentiable in the natural topology. These properties are obvious in the case that both \((U_i, \phi_i)\) and \((U_j, \phi_j)\) belong to either one of the atlases \(\mathcal{A}\) or \(\mathcal{A}''\), but we have to make sure that this is also true for \((U_i, \phi_i) \in \mathcal{A}\) and \((U_j, \phi_j) \in \mathcal{A}''\). Let this be the case, and let \(\mathcal{A}' = \{(U_k, \phi_k), k \in \mathcal{K}\}\). Using the fact that the~\(U_k\) cover \(M_Q\), we may write
\begin{equation}
\begin{split}
\phi_i(U_i \cap U_j) &= \bigcup_{k \in \mathcal{K}}\phi_i(U_i \cap U_j \cap U_k)
= \bigcup_{k \in \mathcal{K}}(\phi_{ik} \circ \phi_k)((U_i \cap U_k) \cap (U_j \cap U_k))\\
&= \bigcup_{k \in \mathcal{K}}\phi_{ki}^{-1}(\phi_k(U_i \cap U_k) \cap \phi_k(U_j \cap U_k))
\end{split}
\end{equation}
Both \(\phi_k(U_i \cap U_k)\) and \(\phi_k(U_j \cap U_k)\) are open, since \(\mathcal{A} \cup \mathcal{A}'\) and \(\mathcal{A}' \cup \mathcal{A}''\) are atlases; hence their intersection is open. Since \(\phi_{ki}\) is continuous, the pre-image of this intersection is open. Finally, since any union of open sets is open, \(\phi_i(U_i \cap U_j)\) is open, which shows property \textit{(iii)}. To see that \(\phi_{ji} = \phi_j \circ \phi_i^{-1}\) is continuous and differentiable in \(\psi \in \phi_i(U_i \cap U_j)\), we choose \(k \in \mathcal{K}\) such that \(\phi_i^{-1}(\psi) \in U_k\) and consider the map
\begin{equation}
\phi_{ji}|_{\phi_i(U_i \cap U_j \cap U_k)} = \phi_{jk}|_{\phi_k(U_i \cap U_j \cap U_k)} \circ \phi_{ki}|_{\phi_i(U_i \cap U_j \cap U_k)}\,.
\end{equation}
Both maps on the right hand side are continuous and differentiable, so the composition is. Using again that the $U_k$ cover $M_Q$ then gives property \textit{(iv)}. So the compatibility of quantum atlases is indeed an equivalence relation. Each equivalence class of compatible quantum atlases of dimension~\(n\) on a set \(M_Q\) provides the structure of a manifold for the underlying set $M_Q$. With these preparations we can now state the central definition:

\vspace{6pt}\noindent\textit{\textbf{Definition.} A quantum manifold of dimension $n$ is a set \(M_Q\) equipped with an equivalence class of quantum atlases of dimension \(n\). The elements of $M_Q$ will be called quantum points.}\vspace{6pt}

In order to avoid confusion we remark that a quantum manifold of dimension $n$ is of course infinite-dimensional. The finite $n$ simply specifies $\mathbb{R}^n$ as the base of the underlying function space. As a topological manifold, a quantum manifold is locally homeomorphic to $\mathcal{S}^{\neq 0}(\mathbb{R}^n)$ with the expectation value topology or the natural topology, which simply follows from the defining properties of a quantum atlas. Since we also required the differentiability of the chart transition functions with respect to the natural topology, the quantum manifold becomes a differentiable manifold as well.

In the following sections we will analyze some important properties of quantum manifolds. In particular, we will consider their classical limit in the next section~\ref{cllimit}; then we prove a nice structural result in section~\ref{fiberbundle}, whereby a quantum manifold is actually a very specific fiber bundle for which the base manifold coincides with the classical limit manifold. Moreover, we show in section~\ref{trivq} that any given classical differentiable manifold can be trivially blown up into a quantum manifold the classical limit of which returns the original manifold.

\section{Classical limit}\label{cllimit}
No construction of a quantum manifold could be useful without a notion of how to reconstruct the differentiable manifold that we interpret as classical spacetime. Starting from the definition of the quantum manifold, we will show in this section how to perform such a classical limit. The basic observation is that not all quantum points are topologically distinguishable in the expectation value topology. Indeed, the indistinguishable quantum points have as their chart images functions with equal position expectation value. We will then show that a suitable Kolmogorov quotient restores the classical structure of a finite-dimensional differentiable manifold with charts essentially provided by the expectation value map.

Since the topology on the quantum manifold $M_Q$ of dimension $n$ lifts from the topology of its model space via the chart homeomorphisms, we first consider the distinguishability of functions $f,g$ in $\mathcal{S}^{\neq 0}(\mathbb{R}^n)$. The relevant topology is the expectation value topology, i.e., the initial topology $\iota (\mathcal{S}^{\neq 0}(\mathbb{R}^n),\bar{\bm{Q}})$ with respect to the position expectation value~$\bar{\bm{Q}}$, as introduced in section~\ref{modelspace}.

Two functions~$f$ and~$g$ are topologically indistinguishable elements of $\mathcal{S}^{\neq 0}(\mathbb{R}^n)$ if and only if the values of their position expectation value vectors coincide, $\bar{\bm{Q}}(f)=\bar{\bm{Q}}(g)$. To prove this we proceed in two steps. First, we show that, if the position expectation values coincide, then every open set $V \in \iota(\mathcal{S}^{\neq 0}(\mathbb{R}^n),\bar{\bm{Q}})$ containing \(f\) also contains \(g\) (and vice versa by the interchange of \(f\) and \(g\)). So let \(f \in V\) and $V$ open. From our constructive characterization of the initial topology we know that there exists an open set \(W \subset \mathbb{R}^n\) so that \(V = \bar{\bm{Q}}^{-1}(W)\) and \(\bar{\bm{Q}}(f) \in W\). Hence \(\bar{\bm{Q}}(g) = \bar{\bm{Q}}(f) \in W\) and \(g \in \bar{\bm{Q}}^{-1}(W) = V\). In the second step of the proof, we consider the case \(\bar{\bm{Q}}(f) \neq \bar{\bm{Q}}(g)\). Then one may choose an open set \(W \subset \mathbb{R}^n\) with \(\bar{\bm{Q}}(f) \in W\) and \(\bar{\bm{Q}}(g) \notin W\), since \(\bar{\bm{Q}}(f)\) and \(\bar{\bm{Q}}(g)\) are distinguishable in the standard topology of \(\mathbb{R}^n\). Thus, \(f \in \bar{\bm{Q}}^{-1}(W)\) and \(g \notin \bar{\bm{Q}}^{-1}(W)\), leading to the conclusion that \(f\) and \(g\) then are topologically distinguishable in the expectation value topology.

Because of the continuity of the chart transition function of the quantum manifold with respect to the expectation value topology the notion of topological distinguishability also lifts from the model space to the quantum manifold: two quantum points can be said to be topologically indistinguishable when their images in some chart are; this statement is chart-independent. Hence it will make sense to apply the Kolmogorov quotient to the quantum manifold.

We now present and prove the main theorem of this section which provides us with a classical limit of the quantum manifold by topological identification of the indistinguishable quantum points.

\vspace{6pt}\noindent\textit{\textbf{Theorem 1.} The Kolmogorov quotient \(M\) of a quantum manifold \(M_Q\) of dimension \(n\) with respect to the expectation value  topology is a differentiable manifold of dimension \(n\) locally homeomorphic to \(\mathbb{R}^n\).}\vspace{3pt}

\noindent\textit{Proof.} We will use two steps in order to prove this theorem. First, we will construct an atlas with the required properties on \(M\) from a quantum atlas on \(M_Q\). Second, we will show that compatible quantum atlases on \(M_Q\) lead to compatible atlases on \(M\).

Let \(\mathcal{A} = \{(U_i, \phi_i), i \in \mathcal{I}\}\) be a quantum atlas of dimension \(n\) on a set \(M_Q\); let \(M\) be the Kolmogorov quotient of \(M_Q\) and \(\mathcal{Q}: M_Q \to M\) the Kolmogorov projection to the equivalence classes of topologically indistinguishable elements. Define subsets \(X_i := \{\mathcal{Q}(\psi), \psi \in U_i\} \subset M\). Then it follows that
\begin{equation}
\mathcal{Q}^{-1}(X_i)=\bigcup_{\psi \in U_i}\mathcal{Q}^{-1}(\mathcal{Q}(\psi))=U_i\,.
\end{equation}
The latter equality holds because the sets \(U_i\) are open in the initial topology on~\(M_Q\); so, for each \(\psi \in U_i\), the equivalence class of elements of \(M_Q\) topologically indistinguishable from \(\psi\), is entirely included in \(U_i\), and \(U_i\) can be written in the form above, as the union of such equivalence classes. From $\mathcal{Q}^{-1}(X_i)$ open we thus conclude that the set \(X_i\) is open in the quotient space topology.

Now consider the image $V_i=\phi_i(U_i)\subset\mathcal{S}^{\neq 0}(\mathbb{R}^n)$ of $U_i$ under a chart. Since $V_i$ is open in the expectation value topology, it can be written as the pre-image of an open set $W_i\subset\mathbb{R}^n$ as $V_i=\bar{\bm{Q}}^{-1}(W_i)$. The sets \(X_i\) and \(W_i\) consist of equivalence classes of topologically indistinguishable elements of \(U_i\) and \(V_i\), respectively. We now use the fact that \(\phi_i: U_i \to V_i\) is a homeomorphism; it follows that \(\phi_i\) is a bijection that maps equivalence classes to equivalence classes. As a consequence, there exists a unique homeomorphism \(\chi_i: X_i \to W_i\), such that the following diagram commutes:
\begin{equation*}
\xymatrix{
M_Q \supset & U_i \ar[r]^{\phi_i} \ar[d]_{\mathcal{Q}} & V_i \ar[d]^{\bar{\bm{Q}}} & \subset \mathcal{S}^{\neq 0}(\mathbb{R}^n) \\
M \supset & X_i \ar[r]_{\chi_i} & W_i & \subset \mathbb{R}^n}
\end{equation*}

We will now show that the collection \(\{(X_i, \chi_i), i \in \mathcal{I}\}\) presents a differentiable atlas on \(M\) by checking the required properties in turn. First we check that the \(X_i\) cover \(M\). Indeed, for any \(\xi \in M\) we may find a \(\psi \in M_Q\) such that \(\mathcal{Q}(\psi) = \xi\). Since the \(U_i\) cover \(M_Q\), there exists \(i \in \mathcal{I}\) such that \(\psi \in U_i\), hence \(\xi \in \mathcal{Q}(U_i) = X_i\). Second, we know that the \(\chi_i\) are homeomorphisms of \(X_i\) onto open subsets \(W_i\) of \(\mathbb{R}^n\). Consequently, for each \(i, j \in \mathcal{I}\) the set \(\chi_i(X_i \cap X_j) \) is open in \(\mathbb{R}^n\), and the transition maps \(\chi_{ji} = \chi_j \circ \chi_i^{-1}: \chi_i(X_i \cap X_j) \to \chi_j(X_i \cap X_j)\) are also homeomorphisms. So far we can conclude that \(\{(X_i, \chi_i), i \in \mathcal{I}\}\) is an atlas of class \(C^0\). Finally, to see that \(\chi_{ji}\) is differentiable, we consider the following diagram:
\begin{equation*}
\xymatrix{\phi_i(U_i \cap U_j) \ar[rr]^{\phi_{ji}} \ar[ddd]^{\bar{\bm{Q}}} & & \phi_j(U_i \cap U_j) \ar[ddd]^{\bar{\bm{Q}}}\\
& U_i \cap U_j \ar[ul]^{\phi_i} \ar[ur]^{\phi_j} \ar[d]^{\mathcal{Q}} & \\
& X_i \cap X_j \ar[dl]^{\chi_i} \ar[dr]^{\chi_j} & \\
\chi_i(X_i \cap X_j) \ar[rr]^{\chi_{ji}} & & \chi_j(X_i \cap X_j)}
\end{equation*}
The upper triangle of this diagram commutes by definition of the transition functions~\(\phi_{ji}\). As we have shown above, there exist unique functions \(\chi_i\) and \(\chi_j\), such that the left hand side and the right hand side of the diagram also commute. Finally, the lower triangle commutes by definition of the transition functions \(\chi_{ji}\). We may thus conclude that the surrounding square of the diagram commutes, so
\begin{equation}
\bar{\bm{Q}} \circ \phi_{ji} = \chi_{ji} \circ \bar{\bm{Q}}\,.
\end{equation}
To solve for the transition function $\chi_{ji}$ we make a convenient (non-unique) choice of an inverse map~$\Psi$ for $\bar{\bm{Q}}$, defined by
\begin{equation}
\Psi :  \mathbb{R}^n  \to  \mathcal{S}^{\neq 0}(\mathbb{R}^n)\,,\,
 \bm{x} \mapsto  \left(\bm{y} \mapsto e^{-(\bm{y} - \bm{x})^2}\right).
\end{equation}
Obviously, \(\Psi(\bm{x})\) is an element of \(\mathcal{S}^{\neq 0}(\mathbb{R}^n)\) and \(\bar{\bm{Q}}(\Psi(\bm{x})) = \bm{x}\) for all \(\bm{x} \in \mathbb{R}^n\), i.e., \(\bar{\bm{Q}} \circ \Psi = \mathrm{id}_{\mathbb{R}^n}\). Composing the equation above with \(\Psi\) from the right, we thus find
\begin{equation}
\bar{\bm{Q}} \circ \phi_{ji} \circ \Psi = \chi_{ji} \circ \bar{\bm{Q}} \circ \Psi = \chi_{ji}
\end{equation}
A quick calculation shows that \(\Psi\) is differentiable with respect to the natural topology on $\mathcal{S}^{\neq 0}(\mathbb{R}^n)$; by definition of the quantum manifold \(\phi_{ji}\) is a diffeomorphism; the expectation value \(\bar{\bm{Q}}\) is differentiable, as we have proven. Hence the composition \(\chi_{ji}\) is differentiable. By a similar argument, we may conclude that also \(\chi_{ji}^{-1} = \chi_{ij}\) is, which shows \(\chi_{ji}\) is a diffeomorphism. This demonstrates that \(\{(X_i, \chi_i), i \in \mathcal{I}\}\) is an atlas of class~\(C^1\).

To complete the proof of the theorem, we have to show that compatible quantum atlases on \(M_Q\) induce compatible atlases on \(M\), i.e., that the manifold structure induced on \(M\) is independent of the choice of an atlas on \(M_Q\). Let \(\mathcal{A}, \mathcal{A}'\) be two compatible quantum atlases on \(M_Q\), inducing atlases \(\tilde{\mathcal{A}}, \tilde{\mathcal{A}}'\) on \(M\). Then, \(\mathcal{A} \cup \mathcal{A}'\) is also a quantum atlas on \(M_Q\). The atlas on \(M\) that is induced by \(\mathcal{A} \cup \mathcal{A}'\) is \(\tilde{\mathcal{A}} \cup \tilde{\mathcal{A}}'\). Since \(\tilde{\mathcal{A}} \cup \tilde{\mathcal{A}}'\) is an atlas, \(\tilde{\mathcal{A}}\) and \(\tilde{\mathcal{A}}'\) are compatible.~$\square$

We further note that on any manifold with a \(C^k\) structure for \(k > 0\) there exists a unique $C^k$-compatible $C^{\infty}$-structure~\cite{Whitney:1936}. Therefore, in the classical limit, we may not only obtain a $C^1$-manifold from our quantum manifold construction, but indeed a $C^{\infty}$-manifold.

\section{Quantum manifolds as fiber bundles}\label{fiberbundle}
In the previous section we have seen how classical spacetime emerges from a quantum manifold and that both are related via the Kolmogorov projection. We will now have a closer look at this construction and investigate further mathematical consequences. It will turn out that the quantum manifold has a structure of particular interest, which is that of a fiber bundle. We will show that the base manifold of this fiber bundle is the classical limit manifold and that the projection onto the base manifold is given by the Kolmogorov projection.

\subsection{Schwartz space as a fiber bundle}\label{Schfib}
We will first show that the model space \(\mathcal{S}^{\neq 0}(\mathbb{R}^n)\) of a quantum manifold, i.e., the Schwartz space with the zero function removed, carries the structure of a trivial fiber bundle with base manifold~\(\mathbb{R}^n\). In the following subsection, we will then use the fact that each chart of a quantum manifold inherits the trivial fiber bundle structure from the Schwartz space to show that also the whole quantum manifold is a fibre bundle.

If \(\mathcal{S}^{\neq 0}(\mathbb{R}^n)\) were a fiber bundle over \(\mathbb{R}^n\), we  might expect that the projection onto the base manifold would be given by the position expectation value \(\bar{\bm{Q}}\). If so, the typical fiber would have to be the pre-image of an arbitrary point in \(\mathbb{R}^n\) under \(\bar{\bm{Q}}\). Without loss of generality, we could choose this point to be the origin. This motivates the definition of the space $\mathcal{S}_0(\mathbb{R}^n)$ of all Schwartz functions with zero position expectation value,
\begin{equation} 
\mathcal{S}_0(\mathbb{R}^n) = \bar{\bm{Q}}^{-1}(0) \subset \mathcal{S}^{\neq 0}(\mathbb{R}^n)\,.
\end{equation}

Functions that do not have expectation value zero can be obtained from elements of $
\mathcal{S}_0(\mathbb{R}^n)$ by means of the translation operator \(T\), which is defined as
\begin{equation}
T :  \mathbb{R}^n \times \mathcal{S}(\mathbb{R}^n)  \to  \mathcal{S}(\mathbb{R}^n)\,,\,
(\bm{x}, f)  \mapsto  T_{\bm{x}}f = (\bm{y} \mapsto f(\bm{y} - \bm{x}))\,.
\end{equation}
Thus, \(T_{\bm{x}}\) translates a function \(f \in \mathcal{S}_0(\mathbb{R}^n)\) by shifting its argument by a constant vector \(\bm{x} \in \mathbb{R}^n\). Clearly, the map \(T_{\bm{x}}: \mathcal{S}_0(\mathbb{R}^n) \to \mathcal{S}_0(\mathbb{R}^n)\) is linear, as one may easily check. We further list some important properties of the translation operator, which will be needed in the following construction. First, we have
\begin{equation}
T_{\bm{y}}T_{\bm{x}}f = T_{\bm{x} + \bm{y}}f
\end{equation}
for all \(\bm{x}, \bm{y} \in \mathbb{R}^n\) and \(f \in \mathcal{S}_0(\mathbb{R}^n)\), 
which immediately follows from the definition of the translation operator. Second, we note that the translation operator \(T_{\bm{x}}\) shifts the position expectation value by \(\bm{x}\), i.e., for all \(\bm{x} \in \mathbb{R}^n\) and \(f \in \mathcal{S}_0(\mathbb{R}^n)\),
\begin{equation}
\bar{\bm{Q}}(T_{\bm{x}}f) = \bar{\bm{Q}}(f) + \bm{x}\,.
\end{equation}
A third important property of the translation operator is its continuity. The map \({T: \mathbb{R}^n \times \mathcal{S}(\mathbb{R}^n) \to \mathcal{S}(\mathbb{R}^n)}\) is continuous with respect to the natural topology on \(\mathcal{S}(\mathbb{R}^n)\) and the corresponding product topology on \(\mathbb{R}^n \times \mathcal{S}(\mathbb{R}^n)\). We give a proof of this in appendix~\ref{proofs}.

With these preparations in place, we are now able to demonstrate that \(\mathcal{S}^{\neq 0}(\mathbb{R}^n)\) is indeed a trivial fiber bundle over \(\mathbb{R}^n\) with typical fiber \(\mathcal{S}_0(\mathbb{R}^n)\), which means there exists a homeomorphism between \(\mathcal{S}^{\neq 0}(\mathbb{R}^n)\) and \(\mathbb{R}^n \times \mathcal{S}_0(\mathbb{R}^n)\). We claim that such a homeomorphism is given by
\begin{equation}\label{taudef}
\tau :  \mathcal{S}^{\neq 0}(\mathbb{R}^n)  \to  \mathbb{R}^n \times \mathcal{S}_0(\mathbb{R}^n)\,,\,
f  \mapsto  (\bar{\bm{Q}}(f), T_{-\bar{\bm{Q}}(f)}f)\,.
\end{equation}

Clearly, for all \(f \in \mathcal{S}^{\neq 0}(\mathbb{R}^n)\) the image \(\tau(f)\) is an element of \(\mathbb{R}^n \times \mathcal{S}_0(\mathbb{R}^n)\) by using the shift of expectation values under the translation operator. It is not difficult to check that the inverse of $\tau$ is given by  \(\tau^{-1}(\bm{x},g) = T_{\bm{x}}g\) for all \((\bm{x},g) \in \mathbb{R}^n \times \mathcal{S}_0(\mathbb{R}^n)\), i.e.,
\begin{equation}\label{tauinv}
\tau^{-1} = T|_{\mathbb{R}^n\times\mathcal{S}_0(\mathbb{R}^n)}\,.
\end{equation} 
Before checking that both $\tau$ and $\tau^{-1}$ are continuous, we need to fix topologies on \(\mathcal{S}^{\neq 0}(\mathbb{R}^n)\) and \(\mathbb{R}^n \times \mathcal{S}_0(\mathbb{R}^n)\). We have already dealt with different topologies on \(\mathcal{S}^{\neq 0}(\mathbb{R}^n)\), namely the natural topology and the initial topology that is induced by the position expectation value. We will show that \(\tau\) is a homeomorphism with respect to both topologies on \(\mathcal{S}^{\neq 0}\), provided we choose the corresponding restriction to \(\mathcal{S}_0(\mathbb{R}^n)\), the standard topology on \(\mathbb{R}^n\) and the product topology on \(\mathbb{R}^n \times \mathcal{S}_0(\mathbb{R}^n)\).

The first (and simpler) case we consider is the expectation value topology on \(\mathcal{S}^{\neq 0}(\mathbb{R}^n)\). Its restriction to \(\mathcal{S}_0(\mathbb{R}^n)\) is the trivial topology, i.e., only the empty set and the space \(\mathcal{S}_0(\mathbb{R}^n)\) itself are open. Consequently, the open subsets of \(\mathbb{R}^n \times \mathcal{S}_0(\mathbb{R}^n)\) are exactly the sets \(W \times \mathcal{S}_0(\mathbb{R}^n)\), where \(W \subset \mathbb{R}^n\) is open. From the definition of \(\tau\) it follows that the pre-images of these open sets under \(\tau\) are the sets \(\bar{\bm{Q}}^{-1}(W)\), which are exactly the open sets in the expectation value topology on \(\mathcal{S}^{\neq 0}(\mathbb{R}^n)\). Thus, \(\tau\) maps open sets to open sets, leading to the conclusion that \(\tau\) is a homeomorphism.

In the second case we consider the natural topology on \(\mathcal{S}^{\neq 0}(\mathbb{R}^n)\) and its restriction to \(\mathcal{S}_0(\mathbb{R}^n)\). To show that \(\tau\) is a homeomorphism with respect to these topologies, we will make use of the fact that the translation operator \(T: \mathbb{R}^n \times \mathcal{S}^{\neq 0}(\mathbb{R}^n) \to \mathcal{S}^{\neq 0}(\mathbb{R}^n)\) is continuous with respect to the natural topology. 
Recall that \(\tau: \mathcal{S}^{\neq 0}(\mathbb{R}^n) \to \mathbb{R}^n \times \mathcal{S}_0(\mathbb{R}^n)\) is continuous with respect to the product topology on \(\mathbb{R}^n \times \mathcal{S}_0(\mathbb{R}^n)\) if and only if the combined maps \(p_1 \circ \tau: \mathcal{S}^{\neq 0}(\mathbb{R}^n) \to \mathbb{R}^n\) and \(p_2 \circ \tau: \mathcal{S}^{\neq 0}(\mathbb{R}^n) \to \mathcal{S}_0(\mathbb{R}^n)\) are continuous. Here, \(p_1\) and \(p_2\) denote the projections onto the first and second factor of \(\mathbb{R}^n \times \mathcal{S}_0(\mathbb{R}^n)\), respectively. The combined maps are shown in the following commuting diagram:

\begin{equation}
\xymatrix@=72pt{ & \mathcal{S}^{\neq 0} \ar[r]^{(-\bar{\bm{Q}}, \mathrm{id}_{\mathcal{S}^{\neq 0}})} \ar@<-1ex>[d]_{\tau} \ar[dl]_{\bar{\bm{Q}}} \ar[dr]^{T_{-\bar{\bm{Q}}}} & \mathbb{R}^n \times \mathcal{S}^{\neq 0} \ar[d]^{T} \\
\mathbb{R}^n & \mathbb{R}^n \times \mathcal{S}_0 \ar[l]^{p_1} \ar[r]_{p_2} \ar@<-1ex>[u]_{T|_{\mathbb{R}^n \times \mathcal{S}_0}} & \mathcal{S}_0}
\end{equation}

The left hand side of the diagram shows \(p_1 \circ \tau = \bar{\bm{Q}}\), which is continuous with respect to the natural topology as we have already shown. The right hand side of the diagram shows the combined map \(p_2 \circ \tau = T_{-\bar{\bm{Q}}}\). To see that this map is also continuous, we decompose it into a combination of two maps, shown in the top right corner of the diagram. The first map assigns to each Schwartz function \(f \in \mathcal{S}^{\neq 0}(\mathbb{R}^n)\) the negative of its position eigenvalue \(-\bar{\bm{Q}}(f)\), along with the function itself. This map is continuous, since both components, which are the negative of the position expectation value and the identity, are continuous. The second map is the translation operator, which is applied to the pair \((-\bar{\bm{Q}}(f), f)\). It is also continuous, leading to the conclusion that also the combined map is continuous. The image of this map has position expectation value \(\bar{\bm{Q}}(f) - \bar{\bm{Q}}(f) = 0\) and is thus an element of \(\mathcal{S}_0(\mathbb{R}^n)\).

Having shown that \(\tau\) is continuous, we still have to show that \(\tau^{-1}\) is also continuous. Recall that~\(\tau^{-1}\) is given by
the translation operator \(T\), restricted to \(\mathbb{R}^n \times \mathcal{S}_0(\mathbb{R}^n)\). Since \(T\) is continuous with respect to the product topology on \(\mathbb{R}^n \times \mathcal{S}^{\neq 0}(\mathbb{R}^n)\), its restriction is continuous with respect to the restricted topology on \(\mathbb{R}^n \times \mathcal{S}_0(\mathbb{R}^n)\), which is again the product topology. Therefore we conclude that \(\tau^{-1}\) is continuous and, thus, \(\tau\) is a homeomorphism with respect to the natural topology.

We can thus conclude that \(\mathcal{S}^{\neq 0}(\mathbb{R}^n)\) is homeomorphic to \(\mathbb{R}^n \times \mathcal{S}_0(\mathbb{R}^n)\) and, consequently, \(\mathcal{S}^{\neq 0}(\mathbb{R}^n)\) is a trivial fiber bundle over \(\mathbb{R}^n\) with typical fiber \(\mathcal{S}_0(\mathbb{R}^n)\). Furthermore, we can conclude that for any open set \(W \subset \mathbb{R}^n\), the pre-image \(\bar{\bm{Q}}^{-1}(W)\) is homeomorphic to \(W \times \mathcal{S}_0(\mathbb{R}^n)\), i.e., a trivial fiber bundle over \(W\) with typical fiber \(\mathcal{S}_0(\mathbb{R}^n)\). Moreover, as shown in appendix~\ref{proofs}, the map~$\tau$ is a diffeomorphism, which makes \(\mathcal{S}^{\neq 0}(\mathbb{R}^n)\) a differentiable fiber bundle.

\subsection{Extension to quantum manifolds}
In the previous section we have shown that every pre-image \(\bar{\bm{Q}}^{-1}(W)\subset \mathcal{S}^{\neq 0}(\mathbb{R}^n)\) of an open set \({W \subset \mathbb{R}^n}\) carries the structure of a trivial fiber bundle over \(W\), with typical fiber \(\mathcal{S}_0\) and fiber bundle projection \(\bar{\bm{Q}}\). Recall that we already used the projection \(\bar{\bm{Q}}: \bar{\bm{Q}}^{-1}(W) \to W\) when we constructed an atlas of the classical manifold from a quantum atlas. We will now employ this construction again to show that the fiber bundle structure of the Schwartz space can be lifted to the quantum manifold.

\vspace{6pt}\noindent\textit{\textbf{Theorem 2.} A quantum manifold \(M_Q\) of dimension \(n\), together with the topology induced by the expectation value topology on \(\mathcal{S}(\mathbb{R}^n)\), carries the structure of a fiber bundle $(M_Q,M,\mathcal{Q})$, where the base manifold \(M\) is the classical limit of \(M_Q\), the projection \(\mathcal{Q}: M_Q \to M\) is given by the Kolmogorov projection \(\mathcal{Q}\), and the typical fiber is the space \(\mathcal{S}_0(\mathbb{R}^n)\).}\vspace{3pt}

\noindent\textit{Proof.} By construction, \(\mathcal{Q}\) is a surjection onto \(M\). We have to check that for all \(\xi \in M\), there exists a neighborhood \(X \subset M\), such that \(\mathcal{Q}^{-1}(X)\) is homeomorphic to \(X \times \mathcal{S}_0(\mathbb{R}^n)\) via a homeomorphism
\begin{equation}
\omega :  \mathcal{Q}^{-1}(X)  \to  X \times \mathcal{S}_0(\mathbb{R}^n)\,,\,
\psi \mapsto  \omega(\psi)=(\omega_1(\psi), \omega_2(\psi))
\end{equation}
which satisfies \(\omega_1 = \mathcal{Q}\). In the following, we will write \(\mathcal{S}_0\) as a shorthand for \(\mathcal{S}_0(\mathbb{R}^n)\).

Let \(\mathcal{A}\) be a quantum atlas and \(\xi \in M\). Then there exists a chart \((U, \phi) \in \mathcal{A}\), inducing a chart \((X, \chi)\) of \(M\) via the construction from the preceding proposition, such that \(\xi \in X\). Let \(\phi(U) = V\), \(\chi(X) = W\) and \(\tau: V \to W \times \mathcal{S}_0\) the homeomorphism introduced in the previous subsection. Let \(\omega: U \to X \times \mathcal{S}_0\) be given by \(\omega = (\chi^{-1} \times \mathrm{id}_{\mathcal{S}_0}) \circ \tau \circ \phi\), such that the upper part of the following diagram commutes:
\begin{equation*}
\xymatrix{U \ar[rrr]^{\phi} \ar[dr]^{\omega} \ar[dd]^{\mathcal{Q}} & & & V \ar[dl]^{\tau} \ar[dd]^{\bar{\bm{Q}}}\\
& X \times \mathcal{S}_0 \ar[r]^{\chi \times \mathrm{id}_{\mathcal{S}_0}} \ar[dl]^{p_1} & W \times \mathcal{S}_0 \ar[dr]^{p_1} & \\
X \ar[rrr]^{\chi} & & & W}
\end{equation*}
Note that \(\chi^{-1} \times \mathrm{id}_{\mathcal{S}_0}\) is a homeomorphism in the product topology as it factorizes into two homeomorphisms. Thus \(\omega\) is a homeomorphism by composition. The right hand side of the diagram commutes because \(V\) is a trivial fiber bundle over \(W\). Obviously, also the lower part of the diagram commutes. Finally the surrounding square of the diagram commutes by the construction of \(\chi\). As a consequence, the left hand side of the diagram commutes, showing that $\omega_1=\mathcal{Q}$, which completes the proof.~$\square$

\section{Trivial quantization}\label{trivq}
In this section we show that every classical, $n$-dimensional manifold \(M\) can be obtained as the Kolmogorov quotient of some quantum manifold \(M_Q\), its trivial quantization. An important mathematical consequence of this is the existence of quantum manifolds.

\subsection{Quantum lift}
Recalling the fact that every quantum manifold of dimension \(n\) is a fiber bundle with typical fiber \(\mathcal{S}_0(\mathbb{R}^n)\) over its Kolmogorov quotient, it seems natural to ask which fiber bundles over a classical manifold carry the structure of a quantum manifold. The most intuitive example one may think of is the trivial $\mathcal{S}_0(\mathbb{R}^n)$-bundle over \(M\), which is the product space \(M \times \mathcal{S}_0(\mathbb{R}^n)\). In the following we will show that this is indeed a quantum manifold, which we will call the trivial quantization of \(M\).

Let \(M\) be an $n$-dimensional differentiable manifold with atlas \(\tilde{\mathcal{A}} = \{(X_i, \chi_i), i \in \mathcal{I}\}\), and define the set \(M_Q = M\times \mathcal{S}_0(\mathbb{R}^n) \). To reveal the quantum manifold structure of \(M_Q\), we have to construct a quantum atlas, which will be denoted by \({\mathcal{A}} = \{(U_i, \phi_i), i \in \mathcal{I}\}\). Recall that the open sets \(U\) corresponding to charts of the quantum atlas are homeomorphic to sets of the form \(X \times \mathcal{S}_0(\mathbb{R}^n)\), where \(X \subset M\) is an open set corresponding to some chart of the classical atlas. We thus set
\begin{equation}\label{Udef}
U_i = X_i \times \mathcal{S}_0(\mathbb{R}^n)\,,
\end{equation}
and we define
\begin{equation}\label{phidef}
\phi_i(\xi, g) = T_{\chi_i(\xi)}g\,.
\end{equation}
We will now show that these definitions indeed provide a quantum atlas \(\mathcal{A}\) on $M_Q=M\times \mathcal{S}_0(\mathbb{R}^n)$ by following the steps \textit{(i)-(iv)} involved in the definition of a quantum atlas in section~\ref{qmfd}.

\textit{(i)} Clearly, the sets $U_i$ are subsets of $M_Q$. They cover $M_Q$ because
\begin{equation}
\bigcup_{i \in \mathcal{I}}U_i = \bigcup_{i \in \mathcal{I}}X_i \times \mathcal{S}_0(\mathbb{R}^n) = \Big(\bigcup_{i \in \mathcal{I}}X_i\Big) \times \mathcal{S}_0(\mathbb{R}^n) = M \times \mathcal{S}_0(\mathbb{R}^n) = M_Q\,,
\end{equation}
using the fact that the sets $X_i$ cover $M$.

\textit{(ii)} We have to show that each \(\phi_i\) is a bijection of \(U_i\) onto a set \(\phi_i(U_i) \subset \mathcal{S}^{\neq 0}(\mathbb{R}^n)\). Recall from the previous section  that for every open set \(W \subset \mathbb{R}^n\) there exists a bijection between the product \(W \times \mathcal{S}_0(\mathbb{R}^n)\) and a subset of \(\mathcal{S}^{\neq 0}(\mathbb{R}^n)\) containing exactly those elements \(f\) for which \(\bar{\bm{Q}}(f) \in W\). This bijection is given by
\begin{equation}
\tau^{-1}|_{W \times \mathcal{S}_0(\mathbb{R}^n)} :  W \times \mathcal{S}_0(\mathbb{R}^n)  \to  \bar{\bm{Q}}^{-1}(W)\,,\,
 (\bm{x}, g)  \mapsto  T_{\bm{x}}g\,.
\end{equation}
Choosing \(W = \chi_i(X_i)\), which is possible since the \(\chi_i(X_i)\) are open subsets of \(\mathbb{R}^n\), we obtain
\begin{equation}
\xymatrix{U_i = & X_i \times \mathcal{S}_0 \ar[d]_{(\chi_i, \mathrm{id}_{\mathcal{S}_0})} \ar[dr]^{\phi_i} & \\
& \chi_i(X_i) \times \mathcal{S}_0 \ar[r]^{\tau^{-1}} & \bar{\bm{Q}}^{-1}(\chi_i(X_i))}
\end{equation}
Using the definition (\ref{phidef}) of $\phi_i$, we realize that the diagram commutes, since
\begin{equation}
\tau^{-1}(\chi_i(\xi), g) = T_{\chi_i(\xi)}g = \phi_i(\xi, g)
\end{equation}
for all \((\xi, g) \in U_i\). Clearly, the maps \((\chi_i, \mathrm{id}_{\mathcal{S}_0})\) and \(\tau^{-1}\) are bijections. Hence \(\phi_i\) is a bijection onto \(\phi_i(U_i) = \bar{\bm{Q}}^{-1}(\chi_i(X_i))\).

\textit{(iii)} In the next step we show that the sets \(\phi_i(U_i \cap U_j)\) are open in the expectation value topology of \(\mathcal{S}^{\neq 0}(\mathbb{R}^n)\). By a similar argument as above, the following diagram commutes:
\begin{equation}
\xymatrix{U_i \cap U_j = & (X_i \cap X_j) \times \mathcal{S}_0 \ar[d]_{(\chi_i, \mathrm{id}_{\mathcal{S}_0})} \ar[dr]^{\phi_i} & \\
& \chi_i(X_i \cap X_j) \times \mathcal{S}_0 \ar[r]^{\tau^{-1}} & \bar{\bm{Q}}^{-1}(\chi_i(X_i \cap X_j))}
\end{equation}
The set \(\chi_i(X_i \cap X_j) \subset \mathbb{R}^n\) is open, since \(\tilde{\mathcal{A}}\) is an atlas of \(M\). Thus, \(\phi_i(U_i \cap U_j) = \bar{\bm{Q}}^{-1}(\chi_i(X_i \cap X_j))\) is the pre-image of an open subset of \(\mathbb{R}^n\) under \(\bar{\bm{Q}}\) and thus open in the expectation value topology.

\textit{(iv)} We show that the transition functions \(\phi_{ji} = \phi_j \circ \phi_i^{-1}\) are continuous with respect to the expectation value topology and differentiable by combining two copies of the preceding diagram:
\begin{equation}
\xymatrix{\phi_i(U_i \cap U_j) \ar[rr]^{\phi_{ji}} & & \phi_j(U_i \cap U_j) \\
& U_i \cap U_j \ar[ul]_{\phi_i} \ar[ur]^{\phi_j} \ar[dl]_{(\chi_i, \mathrm{id}_{\mathcal{S}_0})} \ar[dr]^{(\chi_j, \mathrm{id}_{\mathcal{S}_0})} & \\
\chi_i(X_i \cap X_j) \times \mathcal{S}_0 \ar[uu]^{\tau^{-1}} \ar[rr]_{(\chi_{ji}, \mathrm{id}_{\mathcal{S}_0})} & & \chi_j(X_i \cap X_j) \times \mathcal{S}_0 \ar[uu]_{\tau^{-1}}}
\end{equation}
The left hand side and the right hand side of the diagram commute, as we have already shown. The upper part of the diagram commutes by definition of the transition function \(\phi_{ji}\). The lower part of the diagram commutes by definition of the transition function \(\chi_{ji}\). We can thus conclude that the surrounding square of the diagram also commutes, i.e., that the transition function \(\phi_{ji}\) can be written in the form
\begin{equation}
\phi_{ji} = \tau^{-1} \circ (\chi_{ij}, \mathrm{id}_{\mathcal{S}_0}) \circ \tau
\end{equation}
To see that \(\phi_{ji}\) is continuous with respect to the expectation value topology, we equip the product spaces \(\chi_i(X_i \cap X_j) \times \mathcal{S}_0(\mathbb{R}^n)\) and \(\chi_j(X_i \cap X_j) \times \mathcal{S}_0(\mathbb{R}^n)\) with the product topology, constructed from the standard topology of \(\mathbb{R}^n\) for the first factor and the trivial topology for the second factor. Then,~\(\tau\) is a homeomorphism, i.e., both \(\tau\) and \(\tau^{-1}\) are continuous. Clearly, \((\chi_{ij}, \mathrm{id}_{\mathcal{S}_0})\) is continuous since \(\chi_{ji}\) is continuous with respect to the standard topology of \(\mathbb{R}^n\). Thus, the combined function \(\phi_{ji}\) is also continuous.
Finally, recall that \(\tau\) is not only a homeomorphism, but also a diffeomorphism, i.e. \(\tau\) and \(\tau^{-1}\) are differentiable maps. Furthermore, \((\chi_{ij}, \mathrm{id}_{\mathcal{S}_0})\) is differentiable, since \(\chi_{ji}\) is differentiable by the requirement that \(M\) is a differentiable manifold. This leads to the conclusion that \(\phi_{ji}\) is also differentiable.

This completes the proof that $\mathcal{A}=\{(U_i,\phi_i)\,|\,i\in\mathcal{I}\}$ with $U_i$ and $\phi_i$ defined by equations~(\ref{Udef}) and~(\ref{phidef}) is indeed a quantum atlas of dimension \(n\) on \(M_Q\). Thus, \(M_Q= M\times \mathcal{S}_0(\mathbb{R}^n)\) is an $n$-dimensional quantum manifold. 

\subsection{Classical limit of a trivial quantization}
It remains to be shown that the classical manifold \(M\) can be re-obtained from its trivial quantization \(M_Q= M\times \mathcal{S}_0(\mathbb{R}^n)\) by taking the Kolmogorov quotient and that the classical atlas that corresponds to the quantum atlas $\mathcal{A}$ defined above is the original atlas \(\tilde{\mathcal{A}}\) on $M$.

To construct the Kolmogorov quotient of \(M_Q\), let \((\xi, g)\), \((\xi', g')\) be two topologically indistinguishable elements of \(M\). Further, let \((U_i, \phi_i) \in {\mathcal{A}}\) be a chart of \(M_Q\) with \((\xi, g) \in U_i\), which corresponds to a chart \((X_i, \chi_i) \in \tilde{\mathcal{A}}\). Since \(U_i\) is the pre-image under \(\phi_i\) of a subset of \(\mathcal{S}^{\neq 0}(\mathbb{R}^n)\), which is open in the expectation value topology, \(U_i\) is also open in the induced topology on \(M_Q\). Thus, \((\xi', g') \in U_i\), since topologically indistinguishable elements are contained in the same open sets. Since \(\phi_i\) is a homeomorphism with respect to the induced topology on \(M_Q\), the images \(\phi_i(\xi, g) = T_{\chi_i(\xi)}g\) and \(\phi_i(\xi', g') = T_{\chi_i(\xi')}g'\) are topologically indistinguishable, i.e., their position expectation values \(\bar{\bm{Q}}(T_{\chi_i(\xi)}g) = \chi_i(\xi)\) and \(\bar{\bm{Q}}(T_{\chi_i(\xi')}g') = \chi_i(\xi')\) coincide. This is the case if \(\xi = \xi'\), as \(\chi_i\) is a bijection. The equivalence class \([(\xi, g)]\) of topologically indistinguishable elements thus contains all elements \((\xi', g')\), for which \(\xi = \xi'\), i.e. \([(\xi, g)] = \xi \times \mathcal{S}_0(\mathbb{R}^n)\). This allows us to identify the equivalence classes of topologically indistinguishable elements of \(M_Q\) with elements of \(M\), i.e., writing the Kolmogorov projection as
\begin{equation}
\mathcal{Q} :  M_Q  \to  M\,,\,
 (\xi, g)  \mapsto  \xi\,.
\end{equation}

We still have to check that the topology on \(M\) induced by the Kolmogorov projection is the same as the topology generated by the atlas \(\tilde{\mathcal{A}}\). This means we have to show that any set \(X \subset M\) is open in the topology underlying the atlas $\tilde{\mathcal{A}}$ if and only if it is open in the induced topology (which means its pre-image \(\mathcal{Q}^{-1}(X)\) under the Kolmogorov projection \(\mathcal{Q}\) is open in \(M_Q\)).

The first implication, that open with respect to $\tilde{\mathcal{A}}$ implies open in the induced topology follows from the continuity of $\mathcal{Q}$. Indeed, we can show  that \(\mathcal{Q}\) is continuous on each chart \((U_i, \phi_i)\) of the atlas \(\tilde{\mathcal{A}}\), which in turn proves the continuity of \(\mathcal{Q}\) on the whole quantum manifold \(M_Q\). Recall that within any chart \((U_i, \phi_i)\), the Kolmogorov quotient can be written in the form \(\mathcal{Q} = \chi_i^{-1} \circ \bar{\bm{Q}} \circ \phi_i\). The maps \(\chi_i\) and \(\phi_i\) are homeomorphisms by the definition of the topologies of \(M\) and \(M_Q\). The position expectation value \(\bar{\bm{Q}}\) is continuous by the definition of the expectation value topology on \(\mathcal{S}^{\neq 0}(\mathbb{R}^n)\). Therefore, the composition \(\mathcal{Q}\) is continuous.

To prove the converse, we assume that $X\subset M$ is open in the induced topology, i.e., that $\mathcal{Q}^{-1}(X)$ is open on $M_Q$. We now show that every open set $U\subset M_Q$ can be written as $U=\mathcal{Q}^{-1}(\tilde X)$ for some $\tilde X\subset M$ which is open with respect to $\tilde{\mathcal{A}}$. But then $X=\tilde X$, which proves $X$ is open with respect to $\tilde{\mathcal{A}}$.  So let \(U \subset M_Q\) be open. Then we can use the quantum atlas to write
\begin{equation}
U = \bigcup_{i \in \mathcal{I}}U \cap U_i = \bigcup_{i \in \mathcal{I}}\phi_i^{-1}(\phi_i(U \cap U_i))
\end{equation}
Since \(\phi_i\) is a homeomorphism, the set \(\phi_i(U \cap U_i)\) is open in the expectation value topology of \(\mathcal{S}^{\neq 0}(\mathbb{R}^n)\), i.e. there exist an open set \(W_i^U \subset \mathbb{R}^n\), such that \(\phi_i(U \cap U_i) = \bar{\bm{Q}}^{-1}(W_i^U)\). We can thus rewrite \(U \cap U_i\) as
\begin{eqnarray}
U \cap U_i  & = &  \{(\xi, g) \in U_i\,|\, \bar{\bm{Q}}(\phi_i(\xi, g)) \in W_i^U\} = 
\{(\xi, g) \in U_i\,|\, \chi_i(\xi) \in W_i^U\} \nonumber\\
&= & \chi_i^{-1}(W_i^U) \times \mathcal{S}_0(\mathbb{R}^n) = \mathcal{Q}^{-1}(\chi_i^{-1}(W_i^U))\,,
\end{eqnarray}
using the definition of \(\phi_i\) in the second step. Inserting this equation into the decomposition of \(U\) above leads to
\begin{equation}
U = \bigcup_{i \in \mathcal{I}}\mathcal{Q}^{-1}(\chi_i^{-1}(W_i^U)) = \mathcal{Q}^{-1}\Big(\bigcup_{i \in \mathcal{I}}\chi_i^{-1}(W_i^U)\Big).
\end{equation}
From this equation we see that every open set \(U \subset M_Q\) is the pre-image under \(\mathcal{Q}\) of a union of sets \(\chi_i^{-1}(W_i^U) \subset M\), which are open in the topology generated by the atlas \(\tilde{\mathcal{A}}\). Thus, every open set \(U \subset M_Q\) is the pre-image of an open set \(W \subset M\).

By the results of this section, we have established a one-to-one correspondence between differentiable manifolds and their trivial quantization which is a quantum manifold. So the existence of quantum manifolds is ensured. However, it should be emphasized that quantum manifolds admit a much richer structure than that of a trivial fiber bundle. They allow for non-trivial fibrations as does, in a classical analogy, the M\"obius strip which fibers non-trivially over the circle.  

\section{Discussion}\label{discuss}
Motivated by the ubiquity of function spaces in quantum mechanics and field theory and by the phenomenal success of differentiable manifold geometry in gravity, we have developed a mathematical framework for quantum manifolds in this article. Quantum manifolds are infinite-dimensional manifolds locally homeomorphic to an appropriate space of Schwartz functions, and thus unify geometric formulation and the prominence of function spaces.

Through the topological identification of quantum points, that are charted as functions having the same position expectation value, we have been able to show that there exists a natural way of obtaining a classical limit geometry from the quantum manifold. This procedure is known as the Kolmogorov quotient, and we demonstrated that the classical limit yields nothing else but a finite-dimensional differentiable manifold. The existence of the classical limit inbuilt into our construction is an all-important feature if the mathematical idea of the quantum manifold is to be applied to physical modelling. It is difficult to imagine how any quantum geometry of spacetime could hope for a successful interpretation without clarifying how to reobtain a classical differentiable background.

We could prove that a quantum manifold has the structure of a fiber bundle over its associated classical limit geometry. The projection is the topological identification given by the Kolmogorov projection, and the typical fibers are charted as functions that all map to zero position expectation value.  Making use of their fibre bundle structure we could also show the existence of quantum manifolds. Simple examples are in fact given by finite-dimensional differentiable manifolds to which the typical fibers are all trivially attached. We call this process trivial quantization; the classical limit of a trivial quantization returns the original manifold as one might expect. However, as in the case of the non-trivially fibered M\"obius strip in finite-dimensional differential geometry, trivial fiber bundles cannot capture the full breadth of possible geometries of a quantum manifold. In some sense a quantum manifold is the opposite of a discretized manifold, by blowing up every classical point into a whole fiber of functions.

Though our motivation is physical, this article remains mathematical with the aim to lay the foundations for a rather ambitious project: the proposal to investigate whether physical theories formulated on the infinite-dimensional quantum manifold geometry may appear as quantum theories over a finite-dimensional geometry. Several more ingredients, mathematical as well as physical, will be needed in order to make progress along these lines. Most important on the mathematical side certainly is a clear understanding of tensor bundles over the quantum manifold and their precise relation to the respective bundles over the classical limit geometry. On the physical side this will enable us to start phenomenological model building. Independent of the mathematical interest, the link of quantum manifolds to physics needs be a primary goal for future work.

Ending on a note of speculation, the model building could involve quantum observers given by curves with frames attached on the quantum manifold. These might measure the components of certain tensor fields over the quantum manifold, at least in theory. In practice, any real measurement seems to be restricted to events on a classical manifold. Since this manifold arises by projection as the classical limit, this means that a real observer would only have access to partial information on the observed system. Once the mathematical framework is extended, it will be interesting to investigate whether this would lead to hidden variables problems, or to an appealing  probabilistic or statistical interpretation. We are confident that quantum manifolds will find application since they incorporate what seems to be the most fundamental measurement available: position measurement, not in the absolute, but in the modern relativistic sense, as position coordinates embedded in the differential geometry of the classical limit manifold.

The current status of this research is foundational, and must now pose more questions than answers. This article is a first step towards building a rigorous mathematical background structure for spacetime. The promise of this framework, if field theory on quantum spacetime could be interpreted as quantum theory on classical spacetime, is that a simple gravity theory on the quantum manifold could yield an exciting proposal for a theory of quantum gravity. So this idea could provide a new unified description of field theory and gravity.

\appendix
\section{Topology}\label{topology}
In this appendix we review a few standard topological concepts which are needed for the constructions in this article.

\subsection{Basic notions}
Recall that a topology is one of the most elementary structures on a set of points, where it provides neighborhood relations by specifying the notion of open sets. To be precise, a \textit{topology}~$\mathcal{T}$ on a set $X$ is a system of subsets of $X$, called the \textit{open sets,} which satisfies the following three properties: \textit{(i)}~the empty set $\emptyset$ and the whole set $X$ itself are open, i.e., they are elements of $\mathcal{T}$, \textit{(ii)}~any finite intersection of open sets is open, and, \textit{(iii)}~any, also infinite, union of open sets is open. The pair $(X,\mathcal{T})$ is called a \textit{topological space}.

A given set $X$ may be equipped with different topologies, say with $\mathcal{T}_1$ and $\mathcal{T}_2$. Consider the special case $\mathcal{T}_1\subset\mathcal{T}_2$, i.e., that all open sets of $X$ contained in the topology $\mathcal{T}_1$ are also contained in~$\mathcal{T}_2$; then we call $\mathcal{T}_1$ \textit{coarser} (weaker) than $\mathcal{T}_2$, and $\mathcal{T}_2$ \textit{finer} (stronger) than $\mathcal{T}_1$.

A map $f:X\rightarrow Y$ between two topological spaces $(X,\mathcal{T})$ and $(Y,\mathcal{T}')$ is called \textit{continuous} if and only if the pre-image of any open set is open, i.e., if for all $V\in\mathcal{T'}$ also $f^{-1}(V)\in \mathcal{T}$.

Two points $x, y \in X$ of a topological space $(X, \mathcal{T})$ are called \textit{topologically indistinguishable}, if the open sets containing $x$ are precisely the open sets containing $y$. In other words, it is not possible to find an open set around one of the points that does not also contain the other. It is not difficult to see that topological indistinguishability on a topological space $(X,\mathcal{T})$ is an equivalence relation~$\sim$. It proves useful to consider the equivalence  classes $[x]=\{y\in X | y\sim x\}$ of this relation and their collection.

The \textit{Kolmogorov quotient} of $(X, \mathcal{T})$ is a topological space $(Y, \mathcal{T}')$, where $Y$ is the set of all equivalence classes of topologically indistinguishable elements of $X$ and a subset of $Y$ is open, if and only if its pre-image under the natural surjection $X \to Y,\, x \mapsto [x]$ is open.

Let \(X\) be a set and \(h_i: X \to Y_i, i \in \mathcal{I}\) a family of functions, mapping into topological spaces \((Y_i, \mathcal{T}_i)\). The \emph{initial topology} on \(X\) with respect to \((h_i, i \in \mathcal{I})\) is the topology generated by the pre-images \(h_i^{-1}(V_i)\), where \(V_i \subset Y_i\) is open. It is the coarsest topology on \(X\) such that all functions \(h_i\) are continuous.

Let \(((X_i, \mathcal{T}_i), i \in \mathcal{I})\) be a family of topological spaces, \(X\) their product space and \(\pi_i: X \to X_i\) the projection onto the $i$-th factor. The \emph{product topology} on \(X\) is the initial topology with respect to the family \((\pi_i, i \in \mathcal{I})\) of projections.

Let \((E, \mathcal{T}_E), (B, \mathcal{T}_B), (F, \mathcal{T}_F)\) be topological spaces and \(\pi: E \to B\) a continuous surjection. The space \(E\) is called a \emph{fiber bundle} over \(B\) with typical fiber \(F\) and projection \(\pi\), if for each \(x \in B\) there exists an open set \(U \in \mathcal{T}_B\) with \(x \in U\) and a homeomorphism \(h: \pi^{-1}(U) \to U \times F\), such that the following diagram commutes:
\begin{equation}
\xymatrix{\pi^{-1}(U) \ar[r]^{h} \ar[d]_{\pi} & U \times F \ar[dl]^{p_1} \\
U &}
\end{equation}
Here, \(p_1\) denotes the projection onto the first factor \(U\).

\subsection{Topological vector spaces}
Topological vector spaces are central to the definition of the differentiability of functions. They combine the notions of vector spaces and topological spaces in a compatible way. To be precise, a \textit{topological vector space} \((E, \mathcal{T}_E)\) over a field \(\mathbb{K}\) is a $\mathbb{K}$-vector space $E$ together with a topology \(\mathcal{T}_E\) on \(E\), so that the operations of addition \(+: E \times E \to E\) and multiplication by a scalar \(\cdot: \mathbb{K} \times E \to E\) are continuous. This reflects the intuitive picture that both the translation of open sets by constant vectors and their rescaling by nonzero factors again yield open sets.
 
Given two topological vector spaces $(E, \mathcal{T}_E)$ and $(F, \mathcal{T}_F)$ we are now able to define the notion of differentiability for functions $f:E\rightarrow F$, see~\cite{Lang}. We call \(f\) \emph{differentiable in \(v_0\in E\)}, if it can be locally approximated by a linear function, i.e., if there exists a linear function \(\lambda: E \to F\) and a neighborhood \(V \subset E\) of \(v_0\), such that
\begin{equation}
f(v_0 + v) = f(v_0) + \lambda(v) + \delta(v)
\end{equation}
for all \(v_0 + v \in V\) and \(\delta\) is sufficiently small. \(\lambda\) is called the \emph{derivative} of \(f\) in \(v_0\), and denoted by the differential \(Df(v_0)\).

Of course, we need to clarify what it means for \(\delta\) to be sufficiently small. This is the case when~\(\delta\) is a so-called \emph{tangent to zero}. For \(\tilde{V} \subset E\) an open neighborhood of 0 and \(\delta: \tilde{V} \to F\), the map \(\delta\) is said to be tangent to zero, if for all open neighborhoods \(W \subset F\) of 0 there exists an open set \(V \subset \tilde{V}\),  a positive number \(\epsilon\) and a function \(o: [-\epsilon, \epsilon] \to \mathbb{R}\), such that
\begin{equation}
\lim_{t \to 0}\frac{o(t)}{t} = 0
\end{equation}
and \(\delta(tV) \subset o(t)W\) for all \(t \in [-\epsilon, \epsilon]\). Intuitively, this means that the pre-images of open neighborhoods \(W \subset F\) of 0 under \(\delta\) shrink faster than linearly with the scaling factor of \(W\). 

\section{Technical proofs}\label{proofs}
In this appendix we include a number of proofs needed for the development of the results in the main text. We will be concerned in turn with the continuity and differentiability of the position expectation value map $\bar{\bm Q}$, with properties of the translation map $T$ and with the differentiability of the map~$\tau$ and its inverse between the Schwartz space and its corresponding trivial fibre bundle. As a matter of convenience, some proofs use the nuclear, some the natural, topology on Schwartz space. Since these two are equivalent, as discussed in section~\ref{modelspace}, the claims are true for both. 

\subsection{Continuity of the position expectation value}
\noindent\textit{Claim.} \textit{The position expectation value \(\bar{\bm{Q}}: \mathcal{S}^{\neq 0}(\mathbb{R}^n) \to \mathbb{R}^n\) defined in section~\ref{modelspace} is continuous with respect to the nuclear topology restricted to \(\mathcal{S}^{\neq 0}(\mathbb{R}^n)\) and the standard topology on~$\mathbb{R}^n$.}

\vspace{6pt}\noindent\textit{Proof.} Let \(W \subset \mathbb{R}^n\) be open and \(V = \bar{\bm{Q}}^{-1}(W) \subset \mathcal{S}^{\neq 0}\). To show that \(V\) is open, we will show that for all \(f_0 \in V\) there exists an open set \(\tilde{V}\subset V\) containing $f_0$. Since \(\bar{\bm{Q}}(f_0) \in W\) and \(W\) is open, there exists \(r > 0\) and a corresponding open subset
\begin{equation}
W_{\bar{\bm{Q}}(f_0),r} := \{\bm{x} \in \mathbb{R}^n\,|\, |\bm{x} - \bar{\bm{Q}}(f_0)| < r\} \subset W\,.
\end{equation}
Now, in order to construct the sought-for open set \(\tilde{V} \subset \bar{\bm{Q}}^{-1}(W_{\bar{\bm{Q}}(f_0),r}) \subset V\), consider \(f \in \mathcal{S} \setminus \{-f_0\}\) and estimate
\begin{eqnarray}
|\bar{\bm{Q}}(f_0 + f) - \bar{\bm{Q}}(f_0)| & \leq & 
\frac{\left(|\left<f, \bm{Q}f_0\right>|+|\left<f_0, \bm{Q}f\right>|+|\left<f, \bm{Q}f\right>|\right)\|f_0\|^2}{\|f_0\|^2\|f_0+f\|^2} \nonumber\\
& & {}+\frac{\left(|\left<f_0, f\right>|+|\left<f, f_0\right>|+\|f\|^2\right)|\left<f_0,\bm{Q}f_0\right>|}{\|f_0\|^2\|f_0+f\|^2}\,, 
\end{eqnarray}
using the definition of the expectation value and the triangular inequality.
Choosing \(f\) from a restricted open set of functions
\begin{equation}\label{V1}
V_1 = \Big\{f \in \mathcal{S}\setminus\{-f_0\} \,|\, \|f\|<\frac{1}{2}\|f_0\| \Big\}
\end{equation} 
it follows that $\|f_0 + f\| > \frac{1}{2}\|f_0\|$, so that we can remove $f$ from the denominator of the expression above. We also use the Cauchy-Schwartz inequality to bound all appearing scalar products by norms, and the fact that
\begin{equation}\label{Qffbounds}
\|f\|\leq \|f\|_1\,,\quad \|\bm{Q}f\|\leq \|f\|_1\,,
\end{equation}
where $\|f\|_1^2=\left<f,(\bm{Q}^2+\bm{P}^2+\openone)f\right>$, as defined in~(\ref{pseminorms}). This leads to
\begin{equation}
|\bar{\bm{Q}}(f_0 + f) - \bar{\bm{Q}}(f_0)| \leq
4\|f\|_1 \|f_0\|^{-3} \left[\left(\|\bm{Q}f_0\| + \|f_0\| + \|f\|_1\right)\|f_0\| + \|\bm{Q}f_0\|\left(2\|f_0\| + \|f\|_1\right)\right]\,.
\end{equation}
It is now clear that we can choose $f$ from a further restricted function set $V_2\subset V_1$ for which $\|f\|_1$ is bounded above so that we achieve
\begin{equation}
|\bar{\bm{Q}}(f_0 + f) - \bar{\bm{Q}}(f_0)| < r
\end{equation}
for all \(f \in V_2 = V_1 \cap V_2\), which is open in the nuclear topology as the intersection of open balls. Hence \(\tilde{V} := (V_1 \cap V_2) + f_0\) is open and \(\tilde{V} \subset \bar{\bm{Q}}^{-1}(W_{\bar{\bm{Q}}(f_0),r}) \subset V\). Since \(0 \in V_1 \cap V_2\), we have \(f_0 \in \tilde{V}\). We have thus proven that every \(f_0 \in V\) has an open neighborhood \(\tilde{V} \subset V\) and thus \(V\) is open. So the preimage $V=\bar{\bm Q}^{-1}(W)$ of every open set $W\subset\mathbb{R}^n$ is open, and $\bar{\bm Q}$ continuous.~$\square$

\subsection{Differentiability of the position expectation value}
\noindent\textit{Claim.} \textit{The position expectation value \(\bar{\bm{Q}}: \mathcal{S}^{\neq 0}(\mathbb{R}^n) \to \mathbb{R}^n\) is differentiable with respect to the nuclear topology restricted to \(\mathcal{S}^{\neq 0}(\mathbb{R}^n)\) and the standard topology on~$\mathbb{R}^n$.}

\vspace{6pt}\noindent\textit{Proof.} According to the definition of differentiability in appendix~\ref{topology}, we have to show that $\bar{\bm Q}$ can be linearly approximated, more precisely, that for all \(f_0 \in \mathcal{S}^{\neq 0}\) there exists an open neighborhood \(\tilde{V} \subset \mathcal{S}\) of \(0\) with \(-f_0 \notin \tilde{V}\), a linear function \(D\bar{\bm Q}(f_0): \mathcal{S} \to \mathbb{R}^n\) and a tangent to zero \(\delta: \tilde{V} \to \mathbb{R}^n\) such that for all \(f \in \tilde{V}\)
\begin{equation}
\bar{\bm{Q}}(f_0 + f) = \bar{\bm{Q}}(f_0) + D\bar{\bm Q}(f_0)(f) + \delta(f)\,.
\end{equation}
We first calculate the directional derivative $d/dt|_{t=0} (\bar{\bm{Q}}(f_0 + tf))$ to construct $D\bar{\bm Q}(f_0)$, then we show that the remainder~$\delta$ is indeed tangent to zero. Thus we find
\begin{equation}\label{diffQ}
D\bar{\bm Q}(f_0)(f) = \frac{\left<f_0, \bm{Q}f\right> + \left<f, \bm{Q}f_0\right> - \bar{\bm{Q}}(f_0)(\left<f_0, f\right> + \left<f, f_0\right>)}{\left<f_0, f_0\right>}\,.
\end{equation}

We now show that $\delta$ defined by the two equations above is tangent to zero. In order to do so, we follow similar steps as in the preceding section in obtaining an estimate for $\delta(f)$. Employing the triangular inequality, the Cauchy-Schwartz inequality to bound scalar products by norms, and the bounds~(\ref{Qffbounds}) we arrive at
\begin{equation}
|\delta(f)| \leq 4 \|f\|_1^2 \|f_0\|^{-4} \left(3\|f_0\|^2+7\|f_0\|\|\bm{Q}f_0\|+\|f_0\|\|f\|_1+3\|f\|_1\|\bm{Q}f_0\|\right)
\end{equation}
for all $f$ in the same set $V_1$ as defined by (\ref{V1}). We can now choose $f$ from a further restricted set $V_3\subset V_1$ for which $\|f\|_1$ is bounded above in such a way that
\begin{equation}
|\delta(tf)| < t^2 r 
\end{equation}
for any given positive real $r$ and $|t|<1$. Note that  $V_1\cap V_3$ is open as an intersection of open balls. So $\delta$ maps all functions $tf$ in the open set $V_1\cap V_3$ into a ball $B_{t^2r}(0)\subset \mathbb{R}^n$ of radius $t^2r$ around the origin. Thus we can achieve $\delta(tf)\subset o(t) W$ with $o(t)=t^2$ for any open set $W\subset\mathbb{R}^n$, proving that $\delta$ is tangent to zero, and $\bar{\bm Q}$ differentiable as claimed.~$\square$

\subsection{Continuity of the translation operator}
Aim of this section is to demonstrate the relevant continuity properties of the translation operator $T$ of Schwartz functions which was defined in section~\ref{Schfib}.
 
\vspace{6pt}\noindent\textit{Claim 1.}  \textit{For all \(\bm{x} \in \mathbb{R}^n\), the translation $T_{\bm{x}}: \mathcal{S}(\mathbb{R}^n) \to \mathcal{S}(\mathbb{R}^n),\, f\mapsto T_{\bm{x}}f$ by $\bm{x}$ is a linear homeomorphism with respect to the natural topology.}

\vspace{6pt}\noindent\textit{Proof.} Linearity of \(T_{\bm{x}}\) simply follows from the definition of $T$, since for all $\lambda\in \mathbb{R}$ and $f,g$ in $\mathcal{S}(\mathbb{R}^n)$:
\begin{eqnarray}
T_{\bm{x}}(\lambda f) & = & (\lambda f)(. - \bm{x}) = \lambda f(. - \bm{x}) = \lambda T_{\bm{x}}f\,,\nonumber\\
T_{\bm{x}}(f + g) & = & (f + g)(. - \bm{x}) = f(. - \bm{x}) + g(. - \bm{x}) = T_{\bm{x}}f + T_{\bm{x}}g\,.
\end{eqnarray}

The inverse of \(T_{\bm{x}}\) is given by \(T_{-\bm{x}}\), so it is sufficient to show that \(T_{\bm{x}}\) is continuous and replace \(\bm{x}\) by \(-\bm{x}\) to show that the inverse is also continuous. By definition \(T_{\bm{x}}\) is continuous if for all open sets \(V \subset \mathcal{S}\) the preimage \(T_{-\bm{x}}(V)\) is open, which is equivalent to say that around every $g_0\in T_{-\bm{x}}(V)$ there exists an open neighborhood $\tilde V\subset T_{-\bm{x}}(V)$. This is what we will now show. 

Since $V\subset\mathcal{S}$ is open, there exists an $r>0$ and a finite family of seminorms \(\|.\|_{\bm{\alpha}_j,\bm{\beta}_j}\) for $j = 1, \ldots, m$ so that
\begin{equation}
V' = \bigcap_{j = 1}^{m}V^r_{\bm{\alpha}_j, \bm{\beta}_j}(T_{\bm x}g_0) \subset V\,.
\end{equation}
For $g\in \mathcal{S}$ consider the expression
\begin{equation}
 \|T_{\bm{x}}(g_0 + g) - T_{\bm{x}}g_0\|_{\bm{\alpha}_j,\bm{\beta}_j}  
= \sup_{\bm{y} \in \mathbb{R}^n}\left|(\bm{y}^{\bm{\alpha}_j}D_{\bm{\beta}_j}g(\bm{y}-\bm{x})\right|
= \sup_{\bm{y} \in \mathbb{R}^n}\left|(\bm{x} + \bm{y})^{\bm{\alpha}_j}D_{\bm{\beta}_j}g(\bm{y})\right|\,.
\end{equation}
Expanding $(\bm{x} + \bm{y})^{\bm{\alpha}_j}$, and using the triangular inequality, it is easy to see that this can be bounded by a sum over seminorms of lower length indices,
\begin{equation}
\|T_{\bm{x}}(g_0 + g) - T_{\bm{x}}g_0\|_{\bm{\alpha}_j,\bm{\beta}_j}  
\leq \sum_{|\gamma|\leq|\alpha_j|} C(\bm{\alpha_j},\bm{\gamma},\bm{x})\|g\|_{\bm{\alpha_j}-\bm{\gamma},\bm{\beta_j}}\,,
\end{equation}
with positive coefficients $C(\bm{\alpha_j},\bm{\gamma},\bm{x}) = C'(\bm{\alpha_j},\bm{\gamma})|\bm{x}^\gamma|$. Choosing functions $g$ in the open set 
\begin{equation}
\bigcap_{j=1}^m\bigcap_{|\gamma|\leq|\alpha_j|} V_{\bm{\alpha_j}-\bm{\gamma},\bm{\beta_j}}^{r_j}(0)
\end{equation}
with all $r_j>0$ small enough, we achieve $\|T_{\bm{x}}(g_0 + g) - T_{\bm{x}}g_0\|_{\bm{\alpha}_j,\bm{\beta}_j}<r$, and so $(g_0+g)\in T_{-\bm{x}}(V')$. It follows that the set defined by
\begin{equation}
\tilde V = \bigcap_{j=1}^m\bigcap_{|\gamma|\leq|\alpha_j|} V_{\bm{\alpha_j}-\bm{\gamma},\bm{\beta_j}}^{r_j}(g_0)
\end{equation}
is an open neighborhood of $g_0$ and $\tilde V\subset T_{-\bm{x}}(V')\subset T_{-\bm{x}}(V)$, which completes the proof.~$\square$

\vspace{6pt}\noindent\textit{Claim 2.} \textit{For all \(f \in \mathcal{S}^{\neq 0}(\mathbb{R}^n)\)
 the map \(Tf: \mathbb{R}^n \to \mathcal{S}(\mathbb{R}^n),\,\bm{x}\mapsto T_{\bm x}f\) is a continuous injection with respect to the natural topology on $\mathcal{S}(\mathbb{R}^n)$ and the standard topology on $\mathbb{R}^n$.}

\vspace{6pt}\noindent\textit{Proof.} Obviously $T_{\bm x}f\neq T_{\bm y}f$ for all $\bm{x}\neq\bm{y}$ and $f\in \mathcal{S}^{\neq 0}(\mathbb{R}^n)$; hence $Tf$ is injective. To prove continuity, let \(V \subset \mathcal{S}\) be open. We need to show that the preimage $(Tf)^{-1}(V)$ is open. The empty set is always open, so assume \(\bm{x_0}\in (Tf)^{-1}(V) \neq \emptyset\). Since $V$ is open, there exists an \(r > 0\) and a finite family of seminorms \(\|.\|_{\bm{\alpha}_j,\bm{\beta}_j}\) for \( j = 1, \ldots, m\) such that
\begin{equation}
V' = \bigcap_{j = 1}^{m}V^r_{\bm{\alpha}_j, \bm{\beta}_j}(T_{\bm{x_0}}f) \subset V\,.
\end{equation}
Define $g_j(\bm{y}) = (x_0 + y)^{\bm{\alpha}_j}D^{\bm{\beta}_j}f(\bm{y})$ to rewrite\begin{eqnarray}
\|T_{\bm{x_0} + \bm{x}}(f) - T_{\bm{x_0}}f\|_{\bm{\alpha}_j,\bm{\beta}_j}  & = & \sup_{\bm{y} \in \mathbb{R}^n}|g_j(\bm{y} - \bm{x}) - g_j(\bm{y})| 
\, = \, \sup_{\bm{y} \in \mathbb{R}^n} \Big|\int_{\bm y}^{\bm{y}-\bm{x}} d\tilde{\bm x}\cdot \textrm{grad }g_j(\tilde{\bm x}) \Big| \nonumber\\
& \leq & \sup_{\bm{y} \in \mathbb{R}^n} \Big( |\bm{x}| \sup_{0\leq t\le 1} |\textrm{grad }g_j(\bm{y}-t\bm{x})|\Big)\nonumber\\
& \leq & |\bm{x}|\sup_{\bm{z}\in\mathbb{R}^n}|\textrm{grad }g_j(\bm{z})|\,.
\end{eqnarray}
The right hand side exists, since \(f\) and so the \(g_j\) are Schwartz functions. For any given $f$ we can choose $\bm{x}$ in an open ball $B_{\tilde r}(0)$ of sufficiently small radius $\tilde r$ so that $\|T_{\bm{x_0} + \bm{x}}(f) - T_{\bm{x_0}}f\|_{\bm{\alpha}_j,\bm{\beta}_j}<r$ for all $j$. Then $(\bm{x}_0+\bm{x})\in (Tf)^{-1}(V')$. It follows that the open neighborhood $B_{\tilde r}(\bm{x}_0)$ of $\bm{x}_0$ satisfies $B_{\tilde r}(\bm{x}_0) \subset (Tf)^{-1}(V') \subset (Tf)^{-1}(V)$. So we conclude that \((Tf)^{-1}(V)\) is open.~$\square$

\vspace{6pt}\noindent\textit{Claim 3.} \textit{The translation operator $T: \mathbb{R}^n \times \mathcal{S}(\mathbb{R}^n)  \to  \mathcal{S}(\mathbb{R}^n),\,(\bm{x}, f)  \mapsto  T({\bm{x}},f)=T_{\bm{x}}f$  is continuous with respect to the natural topology on \(\mathcal{S}(\mathbb{R}^n)\) and the corresponding product topology on \({\mathbb{R}^n \times \mathcal{S}(\mathbb{R}^n)}\).}

\vspace{6pt}\noindent\textit{Proof.} Let \(V \subset \mathcal{S}\) be open and \((\bm{x_0}, f_0) \in T^{-1}(V)\). We will show that there exists an open neighborhood \(Y \subset \mathbb{R}^n \times \mathcal{S}\) of \((\bm{x_0}, f_0)\) such that \(Y \subset T^{-1}(V)\), which proves that the preimage \(T^{-1}(V)\) is open. Since \(V\) is open in the natural topology there exists $r>0$  and a family of seminorms \(\|.\|_{\bm{\alpha}_j,\bm{\beta}_j}\) for \(j = 1, \ldots, m\) so that
\begin{equation}
V' = \bigcap_{j = 1}^{m}V^r_{\bm{\alpha}_j,\bm{\beta}_j}(T({\bm{x_0}},f_0)) \subset V\,.
\end{equation}
Consider for \(\bm{x} \in \mathbb{R}^n\) and \(f \in \mathcal{S}\) the expression
\begin{eqnarray}
\|T(\bm{x}_0+\bm{x},f_0+f)-T(\bm{x}_0,f_0)\|_{\bm{\alpha}_j,\bm{\beta}_j}
& \leq & \|T(\bm{x}_0+\bm{x},f_0+f)-T(\bm{x}_0+\bm{x},f_0)\|_{\bm{\alpha}_j,\bm{\beta}_j} \nonumber\\
& & {} + \|T(\bm{x}_0+\bm{x},f_0)-T(\bm{x}_0,f_0)\|_{\bm{\alpha}_j,\bm{\beta}_j}\,.
\end{eqnarray}
The continuity of $Tf_0$ proven in \textit{Claim 2} of this section tells us that we may find $\tilde r>0$, depending only on~$\bm{x}_0$ and~$f_0$, so that for $\bm{x}\in B_{\tilde r}(0)$ the second term becomes smaller than $r/2$ for all $j$. A closer look at the proof of \textit{Claim 1} in this section reveals that the first term can be made smaller than $r/2$ for all~$j$ by choosing~$f$ in a sufficiently small open neighborhood of $0$ given as a finite intersection of the form $\bigcap V^{r_k}_{\bm{\alpha}_k,\bm{\beta}_k}(0)$, where the $r_k$ again only depend on $\bm{x}_0$ and $f_0$. Combining these facts we obtain $(\bm{x}_0+\bm{x},f_0+f)\in T^{-1}(V')$ for $(x,f)\in B_{\tilde r}(0)\times \bigcap V^{r_k}_{\bm{\alpha}_k,\bm{\beta}_k}(0)$ which is open in the product topology. It follows that 
\begin{equation}
Y= \Big(B_{\tilde r}(\bm{x}_0)\times \bigcap V^{r_k}_{\bm{\alpha}_k,\bm{\beta}_k}(f_0)\Big) \subset T^{-1}(V')\subset T^{-1}(V)
\end{equation}
is an open neighborhood, in the product topology, of $(\bm{x}_0,f_0)$, which completes the proof.~$\square$

\subsection{Differentiability of $\tau$ and $\tau^{-1}$}
We have shown in section \ref{Schfib} that the map $\tau:\mathcal{S}^{\neq 0}(\mathbb{R}^n)\rightarrow \mathbb{R}^n\times \mathcal{S}_0(\mathbb{R}^n)$ as defined in~(\ref{taudef}) is a homeomorphism, making the model space $\mathcal{S}^{\neq 0}(\mathbb{R}^n)$ of the quantum manifold a trivial fibre bundle. Here we will show that $\tau$ is even a diffeomorphism. 

\vspace{6pt}\noindent\textit{Claim 1.} \textit{The map $\tau$ is differentiable, with the natural topology on $\mathcal{S}(\mathbb{R}^n)$. For all $f_0\in \mathcal{S}^{\neq 0}(\mathbb{R}^n)$ the differential $D\tau(f_0)$ of $\tau$ at $f_0$ is given by the linear map}
\begin{eqnarray}
D\tau(f_0) :\mathcal{S}(\mathbb{R}^n) & \rightarrow & \mathbb{R}^n \times \mathcal{S}(\mathbb{R}^n)\,,\nonumber\\
 g & \mapsto & \left(D\bar{\bm{Q}}(f_0)(g),\, D\bar{\bm{Q}}(f_0)(g) \cdot \textrm{grad }(T_{-\bar{\bm{Q}}(f_0)}f_0) + T_{-\bar{\bm{Q}}(f_0)}g\right),
\end{eqnarray}
\textit{where the differential of $\bar{\bm Q}$ is displayed in equation~(\ref{diffQ}).}

\vspace{6pt}\noindent\textit{Proof.} To prove this, we need to show that $\tau$ can be linearly approximated as
\begin{equation}
\tau(f_0 + f) = \tau(f_0) + D\tau(f_0)(f) + \delta(f)
\end{equation}
for all $f$ in a small open neighborhood \({V} \subset \mathcal{S}\) of \(0\), and a tangent to zero $\delta$ defined on $V$. We decompose $\delta(f) = (\delta_1(f), \delta_2(f)) \in \mathbb{R}^n \times \mathcal{S}$ which gives
\begin{align}
\delta_1(f) &= \bar{\bm{Q}}(f_0 + f) - \bar{\bm{Q}}(f_0) - D\bar{\bm{Q}}(f_0)(f)\,,\label{delta1}\\
\delta_2(f) &= T_{-\bar{\bm{Q}}(f_0 + f)}(f_0 + f) - T_{-\bar{\bm{Q}}(f_0)}f_0 - D\bar{\bm{Q}}(f_0)(f) \cdot \textrm{grad }T_{-\bar{\bm{Q}}(f_0)}f_0 - T_{-\bar{\bm{Q}}(f_0)}f\,.
\end{align}
We already know from the proof of the differentiability of $\bar{\bm Q}$ that \(\delta_1\) is a tangent to zero. To show that  $\delta$ is a tangent to zero, it remains to show that \(\delta_2\) is. To do so, we expand \(\delta_2\) as
\begin{eqnarray}
\delta_2(f)(\bm{x}) & = & f_0(\bm{x} + \bar{\bm{Q}}(f_0 + f)) - f_0(\bm{x} + \bar{\bm{Q}}(f_0)) - D\bar{\bm{Q}}(f_0)(f) \cdot \textrm{grad } f_0(\bm{x} + \bar{\bm{Q}}(f_0)) \nonumber\\
& & {}+ f(\bm{x} + \bar{\bm{Q}}(f_0 + f)) - f(\bm{x} + \bar{\bm{Q}}(f_0))\,.
\end{eqnarray}
Then we replace all occurrences of \(\bar{\bm{Q}}(f_0 + f)\) using equation~(\ref{delta1}) and apply Taylor's theorem to the first term in each line, which yields
\begin{eqnarray}\label{deltaexp}
& & x^{\bm{\alpha}}D_{\bm \beta}\delta_2(f)(\bm{x}) \, = \,  \delta_1(f) \cdot \textrm{grad } x^{\bm{\alpha}}D_{\bm \beta}f_0(\bm{x} + \bar{\bm{Q}}(f_0)) + \sum_{|\bm{\gamma}| = 2}R^{x^{\bm{\alpha}}D_{\bm \beta}f_0}_{\bm{\gamma}}(D\bar{\bm{Q}}(f_0)(f) + \delta_1(f))^{\bm{\gamma}}\\
& & \qquad\qquad{} + (D\bar{\bm{Q}}(f_0)(f) + \delta_1(f)) \cdot \textrm{grad } x^{\bm{\alpha}}D_{\bm \beta}f(\bm{x} + \bar{\bm{Q}}(f_0)) + \sum_{|\bm{\gamma}| = 2}R^{x^{\bm{\alpha}}D_{\bm \beta}f}_{\bm{\gamma}}(D\bar{\bm{Q}}(f_0)(f) + \delta_1(f))^{\bm{\gamma}}\nonumber
\end{eqnarray}
where the sums involve only multiindices $\bm{\gamma}$ of length two, and the precise form of the remainder terms $R^{x^{\bm{\alpha}}D_{\bm \beta}f_0}_{\bm{\gamma}}$ and $R^{x^{\bm{\alpha}}D_{\bm \beta}f}_{\bm{\gamma}}$ will not be required. Now choose an open neighborhood $Z\subset \mathcal{S}$ of~$0$. According to the definition of the natural topology on $\mathcal{S}$ there exists a finite family \(\|.\|_{\bm{\alpha}_j,\bm{\beta}_j}\) for \(j = 1, \ldots, m\) and \(r > 0\) so that
\begin{equation}
Z' = \bigcap_{j = 1}^{m}V^r_{\bm{\alpha}_j,\bm{\beta}_j}(0) \subset Z\,.
\end{equation}
For $|t|\leq 1$ consider the expression
\begin{equation}
\|\delta_{2}(tf)\|_{\bm{\alpha}_j,\bm{\beta}_j} = \sup_{\bm{x} \in \mathbb{R}^n}\left|x^{\bm{\alpha}_j}D_{\bm{\beta}_j}\delta_{2}(tf)(\bm{x})\right|\,.
\end{equation}
Using the expansion of $x^{\bm{\alpha}_j}D_{\bm \beta_j}\delta_2(tf)(\bm{x})$ provided by equation~(\ref{deltaexp}) and the estimate $|R^h_{\bm{\gamma}}|\leq \sup_{\bm{y}\in\mathbb{R}^n}|D_{\bm{\gamma}}h(\bm{y})/\bm{\gamma} !|$ for the remainder term in the Taylor series of a Schwartz function $h$, we are then able to show that a bound of the form 
\begin{equation}
\|\delta_{2}(tf)\|_{\bm{\alpha}_j,\bm{\beta}_j} \leq |t|^2 r_j(f)
\end{equation}
holds. We do not display the rather complicated expression $r_j(f)>0$; what matters is the fact that it can be expressed in terms of the seminorms that generate the topology on \(\mathcal{S}\). Moreover, $r_j(f)>0$ can be made arbitrarily small by choosing $f$ from a correspondingly small neighborhood $\tilde V\subset\mathcal{S}$ of $0$ which is open in the natural topology. Hence, choosing $\tilde V$ small enough so that $\|\delta_{2}(tf)\|_{\bm{\alpha}_j,\bm{\beta}_j} \leq |t|^2 r$ for all $j$, it now follows that $\delta_2(t\tilde V)\subset t^2\delta_2(Z')\subset t^2\delta_2(Z)$. So $\delta_2$ is a tangent to zero for $o(t)=t^2$, which completes the proof.~$\square$

\vspace{6pt}We now consider the differentiability of the inverse map $\tau^{-1}:\mathbb{R}^n\times\mathcal{S}_0(\mathbb{R}^n)\rightarrow \mathcal{S}^{\neq 0}(\mathbb{R}^n)$ which, as we know from~(\ref{tauinv}), is given by $\tau^{-1}=T|_{\mathbb{R}^n\times\mathcal{S}_0(\mathbb{R}^n)}$.

\vspace{6pt}\noindent\textit{Claim 2.} \textit{The inverse map $\tau^{-1}$ is differentiable, with the natural topology on $\mathcal{S}(\mathbb{R}^n)$. For all $\bm{x}_0\in\mathbb{R}^n$ and $g_0\in \mathcal{S}_0(\mathbb{R}^n)$ the differential $D\tau^{-1}(\bm{x}_0,g_0)$ of $\tau^{-1}$ at $(\bm{x}_0,g_0)$ is given by the linear map}
\begin{equation}
D\tau^{-1}(\bm{x}_0,g_0) : \mathbb{R}^n \times \left\{g \in \mathcal{S}\,|\, D\bar{\bm Q}(g_0)(g)= 0\right\}  \to  \mathcal{S}\,,\, (\bm{x}, g)  \mapsto  -\bm{x} \cdot \textrm{grad }T_{\bm{x}_0}g_0 + T_{\bm{x}_0}g\,.
\end{equation}

\vspace{6pt}\noindent\textit{Proof.} This proof proceeds similarly as the last one. We need to show that $\tau^{-1}$ can be linearly approximated as
\begin{equation}
\tau^{-1}(\bm{x}_0 + \bm{x}, g_0 + g) = \tau^{-1}(\bm{x}_0, g_0) + D\tau^{-1}(\bm{x}_0, g_0)(\bm{x}, g) + \delta(\bm{x}, g)
\end{equation}
for all $(\bm{x}, g)$ in a sufficiently small open neighborhood $V\subset (\mathbb{R}^n\times\mathcal{S})$ of \((\bm{0},0)\), and a tangent to zero $\delta$ defined on~$V$. From the definitions of $\tau^{-1}$ and $T$, we find
\begin{eqnarray}
\delta(\bm{x}, g)(\bm{y}) 
& = & g_0(\bm{y} - \bm{x}_0 - \bm{x}) - g_0(\bm{y} - \bm{x}_0) + \bm{x} \cdot \textrm{grad } g_0(\bm{y} - \bm{x}_0) \nonumber\\
& & {} + g(\bm{y} - \bm{x}_0 - \bm{x}) - g(\bm{y} - \bm{x}_0)\,.
\end{eqnarray}
Application of Taylor's theorem to the first term in each line yields
\begin{equation}
y^{\bm{\alpha}}D_{\bm{\beta}}\delta(\bm{x}, g)(\bm{y}) = 
\sum_{|\bm{\gamma}| = 2}\tilde R_{\bm{\gamma}}^{y^{\bm{\alpha}}D_{\bm{\beta}}g_0}x^{\bm{\gamma}}
-\bm{x} \cdot \textrm{grad } y^{\bm{\alpha}}D_{\bm{\beta}}g(\bm{y} - \bm{x}_0) + \sum_{|\bm{\gamma}| = 2}\tilde R_{\bm{\gamma}}^{y^{\bm{\alpha}}D_{\bm{\beta}}g}x^{\bm{\gamma}}
\end{equation}
where, as above, the sums involve only multiindices $\bm{\gamma}$ of length two, and the precise form of the remainder terms $\tilde R^{x^{\bm{\alpha}}D_{\bm \beta}f_0}_{\bm{\gamma}}$ and $\tilde R^{x^{\bm{\alpha}}D_{\bm \beta}f}_{\bm{\gamma}}$ will not be required. Now let $S\subset \mathcal{S}$ be an open neighborhood of \(0\). There exists $r>0$ and a finite family of seminorms \(\|.\|_{\bm{\alpha}_j,\bm{\beta}_j}\) for \(j = 1, \ldots, m\) so that 
\begin{equation}
S' = \bigcap_{j = 1}^{m}V^r_{\bm{\alpha}_j,\bm{\beta}_j}(0) \subset {S}\,.
\end{equation}
We consider the expression
\begin{equation}
\|\delta(t\bm{x}, tg)\|_{\bm{\alpha}_j,\bm{\beta}_j} = \sup_{\bm{y} \in \mathbb{R}^n}\left|y^{\bm{\alpha}_j}D_{\bm{\beta}_j}\delta(t\bm{x}, tg)(\bm{y})\right|
\end{equation}
and employ the expansion of $y^{\bm{\alpha}}D_{\bm{\beta}}\delta(\bm{x}, g)(\bm{y})$ obtained above to show that the following bound holds for $\|t\|\leq 1$:
\begin{eqnarray}
\|\delta(t\bm{x}, tg)\|_{\bm{\alpha}_j,\bm{\beta}_j} & \leq &
 |t|^2 \Big(\|\bm{x}\| \sup_{\bm{y} \in \mathbb{R}^n}\|\textrm{grad }y^{\bm{\alpha}_j}\bm{\beta}_j g(\bm{y}-\bm{x}_0)\|\Big. \nonumber\\
& & \qquad\Big.+\|\bm{x}\|^2 \sum_{|\bm{\gamma}| = 2}\sup_{\bm{z} \in \mathbb{R}^n}\frac{1}{\bm{\gamma}!} \left(|D_{\bm{\gamma}}(y^{\bm{\alpha}_j}\bm{\beta}_jg_0)|+|D_{\bm{\gamma}}(y^{\bm{\alpha}_j}\bm{\beta}_jg)|   \right) \Big). 
\end{eqnarray}
It is clear that the term in brackets can be made smaller than $r$ for all $j$ by choosing 
$(\bm{x},g)$ from a sufficiently small set $\tilde V\subset \mathbb{R}^n\times\mathcal{S}$, open in the relevant product topology. It follows that \(\delta(t\tilde V) \subset t^2 S' \subset t^2 {S}\), which shows that $\delta$ is a tangent to zero for $o(t)=t^2$, and hence~$\tau^{-1}$ differentiable as claimed. For completeness, it is not hard to check that the differentials $D\tau$ and~$D\tau^{-1}$ are inverses to one another.~$\square$

\acknowledgments
MH gratefully acknowledges a research scholarship from the Graduiertenkolleg 602. RP and MNRW gratefully acknowledge full financial support from the German Research Foundation DFG through the Emmy Noether fellowship grant WO 1447/1-1.



\end{document}